\documentclass[12pt]{article}
\usepackage{epsfig,cite,amsmath,axodraw,hhline,color}

\input paperdef
\graphicspath{{figs/}}

\evensidemargin \oddsidemargin
\makeatletter
\renewcommand{\section}{\@startsection {section}{1}{\z@}%
                           {-3.5ex \@plus -1ex \@minus -.2ex}%
                           {2.3ex \@plus.2ex}%
                           {\mathversion{bold}\normalfont\Large\bfseries}}
\renewcommand{\subsection}{\@startsection{subsection}{2}{\z@}%
                           {-3.25ex\@plus -1ex \@minus -.2ex}%
                           {1.5ex \@plus .2ex}%
                           {\mathversion{bold}\normalfont\large\bfseries}}
\renewcommand{\subsubsection}{\@startsection{subsubsection}{3}{\z@}%
                           {-3.25ex\@plus -1ex \@minus -.2ex}%
                           {1.5ex \@plus .2ex}%
                           {\mathversion{bold}\normalfont\normalsize\bfseries}}
\makeatother

\newcommand{\pdrei}{{\bf P3}}
\newcommand{\pvier}{{\bf P4}}

\allowdisplaybreaks

\begin{document}
\thispagestyle{empty}

\def\thefootnote{\fnsymbol{footnote}}

\begin{flushright}
DCPT/10/162\\
DESY 10-217\\
IPPP/10/81 \\
arXiv:1012.5007 [hep-ph]
\end{flushright}

\vspace{0cm}

\begin{center}

{\large\sc 
{\bf BSM Higgs Physics in the Exclusive Forward Proton Mode\\[.5em]
at the LHC}}

\vspace{0.0cm}

\vspace{1cm}

{\sc 
S.~Heinemeyer$^{1}$%
\footnote{email: Sven.Heinemeyer@cern.ch}%
, V.A.~Khoze$^{2, 3}$%
\footnote{email: V.A.Khoze@durham.ac.uk}%
, M.G.~Ryskin$^{2, 4}$%
\footnote{email: misha.ryskin@durham.ac.uk}%
,\\[2mm] M.~Tasevsky$^{5}$%
\footnote{email: Marek.Tasevsky@cern.ch}%
~and G.~Weiglein$^{6}$%
\footnote{email: Georg.Weiglein@desy.de}
}

\vspace*{0.5cm}

{\sl
$^1$Instituto de F\'isica de Cantabria (CSIC-UC), 
Santander, Spain

\vspace*{0.25cm} 

$^2$IPPP, Department of Physics, Durham University, 
Durham DH1 3LE, U.K.

\vspace*{0.25cm} 

$^3$School of Physics \& Astronomy, University of Manchester, 
Manchester M13 9PL, U.K.

\vspace*{0.25cm} 

$^4$Petersburg Nuclear Physics Institute, Gatchina, 
St.~Petersburg, 188300, Russia

\vspace*{0.25cm} 

$^5$Institute of Physics, 
18221 Prague 8, Czech Republic

\vspace*{0.25cm}

$^6$DESY, Notkestra\ss e 85, D--22603 Hamburg, Germany

}

\end{center}

\vspace*{0.2cm}
\begin{abstract}
We investigate the prospects for Central Exclusive Diffractive (CED)
production of BSM Higgs bosons  at the LHC using
forward proton detectors installed
at 220~m and 420~m distance around ATLAS and / or CMS.
We update a previous analysis for the MSSM taking into account
improvements in the theoretical calculations and the most
recent exclusion bounds from the Tevatron.
We extend the MSSM analysis to new benchmark scenarios that are in
agreement with the cold dark matter relic abundance and other precision
measurements.
We analyse the exclusive production of Higgs bosons in a model with a
fourth generation of fermions. 
Finally, we comment on the determination of Higgs spin--parity and coupling
structures at the LHC and show that the forward proton mode
could provide crucial information on the $\cp$ properties
of the Higgs bosons.
\end{abstract}

\def\thefootnote{\arabic{footnote}}
\setcounter{page}{0}
\setcounter{footnote}{0}

\newpage

%%%%%%%%%%%%%%%%%%%%%%%%%%%%%%%%%%%%%%%%%%%%%%%%%%%%%%%%%%%%%%%%%%%%%%%%%%%%%%%
%%%%%%%%%%%%%%%%%%%%%%%%%%%%%%%%%%%%%%%%%%%%%%%%%%%%%%%%%%%%%%%%%%%%%%%%%%%%%%%

\section{Introduction}
\label{sec:intro}

There has been a growing interest in the recent years
in the possibility to complement the standard LHC physics menu by 
forward and diffraction physics. This requires the installation of
near-beam proton detectors in the LHC tunnel, see
for example \citeres{ar,KMRProsp,acf,DKMOR,KMRbsm}.
Within ATLAS and CMS
projects to install the proton detectors at 220~m and 420~m from the
interaction points have been discussed~\cite{CMS-Totem,FP420,AFP}.
The combined detection of the centrally produced system and
both outgoing protons gives access to a rich programme
of studies of QCD, electroweak and BSM physics, see for
instance \citeres{FP420,KMRProsp,cr,kp1,bcfp,mt1}. Importantly,
these measurements will
provide valuable information on the Higgs sector of MSSM and other popular
BSM scenarios~\cite{KKMRext,diffH,CLP,fghpp, ismd,tripl,eds09}.

As it is well known, many models of new physics require an extended
Higgs sector. The most popular extension of the SM 
is the MSSM~\cite{susy}, where the Higgs sector consists of
five physical states. At lowest order the MSSM Higgs
sector is $\cp$-conserving, containing two $\cp$-even
bosons, the lighter $h$ and the heavier $H$, 
a $\cp$-odd boson, $A$, and the charged
bosons $H^\pm$. It can be specified
in terms of the gauge couplings, the ratio of the two vacuum
expectation values, $\tb \equiv v_2/v_1$, and the mass of the $A$
boson, $\MA$.
The Higgs sector of the MSSM is affected by large higher-order
corrections (see for example\ \citere{reviews} for recent reviews), which
have to be taken into account for reliable phenomenological predictions.

Another  very simple example of physics beyond the SM is a
model which extends the SM by a fourth generation of heavy fermions
(SM4), see for
instance~\citeres{extra-gen-review,4G,4G-ew,four-gen-and-Higgs,oncemore4G}.  
Here the masses of the 4th generation quarks
are assumed to be (much) heavier than the mass of the top-quark.
In this case, the effective coupling of the Higgs boson
to two gluons is three times larger than in the SM, and all branching
ratios change correspondingly.

Proving that a detected new state is, indeed, a Higgs boson and
distinguishing the Higgs boson(s) of the SM, the SM4 or the MSSM from
the states of other theories will be far from trivial.
In particular, it will be of utmost importance 
to determine the spin and
$\cp$ properties of a new state and to measure precisely its mass, width
and couplings.

Forward proton detectors
installed at 220~m and 420~m  around ATLAS and / or CMS would provide a
rich complementary physics potential to the  
 ``conventional'' LHC Higgs production channels.
The CED processes are of the form
\begin{align}
pp\to p \oplus H \oplus p~, 
\end{align}
where the $\oplus$ signs denote
large rapidity gaps on either side of the   centrally produced state.
If the outgoing protons remain intact and scatter
through small angles then, to a very good approximation, the primary
di-gluon system obeys a $J_z=0$, $\cC$-even, $\cP$-even selection
rule~\cite{KMRmm}. 
Here $J_z$ is the projection of the total
angular momentum along the proton beam axis. This 
permits a clean determination of the quantum numbers of the
observed resonance which  will be dominantly produced in a $0^{++}$ 
state. Furthermore, due to the exclusive nature of the process, the
proton  energy losses are directly related to the central mass. 
This property together with precise tracking detectors provides a
potentially excellent mass resolution 
irrespective of the decay channel (and, depending on the Higgs mass, 
could also provide access to a determination of
the Higgs boson width). The CED
processes allow in principle all the main
Higgs decay modes, $b \bar b$,  $WW$  and $\tau\tau$, to be
observed in the same production channel. In particular, a unique
possibility opens up to study the Higgs Yukawa coupling to bottom
quarks, which, as it is well known, may be difficult
to access in other search channels at the LHC. Here it should be kept in
mind that access to the bottom Yukawa coupling will be crucial as an
input also for the determination of 
Higgs couplings to other particles~\cite{HcoupLHCSM,HcoupLHC120}.

Within the MSSM, CED Higgs production is even more appealing than in the SM. 
The lightest MSSM Higgs-boson coupling to $b \bar b$ and $\tau\tau$ 
can be strongly enhanced for large values of
$\tb$ and relatively small $\MA$. On the other hand, for
larger values of $\MA$ the branching ratio of $H \to b \bar b$ 
is much larger
than for a SM Higgs of the same mass. As a consequence, CED 
$H \to b \bar b$ production can be studied in the MSSM up
to much higher masses than in the SM case.

The outline of the paper is as follows.
We revisit the analysis of~\citere{diffH}
where a detailed study of the CED MSSM Higgs
production was performed (see also \citeres{KKMRext,kmrcp,je1,je2,CLP,cr1} for
other CED studies in the MSSM).
This is updated  by taking into account
recent theoretical developments in background evaluation~\cite{shuv,screen}
and using an improved version of the code
{\tt FeynHiggs}~\cite{feynhiggs,mhiggslong,mhiggsAEC,mhcMSSMlong}
(available from \citere{fh-web})
employed for the cross section and
decay width calculations. The regions excluded by LEP and Tevatron Higgs
searches are evaluated with {\tt HiggsBounds}~\cite{higgsbounds}.
These improvements are applied for the CED production of MSSM Higgs
bosons~\cite{diffH} in the $\Mhmax$ and no-mixing 
benchmark scenarios (defined in \cite{benchmark2}) as well as for new
benchmark scenarios that yield the correct amount of the cold dark
matter abundance.
In \refse{sec:SM4} we show the corresponding results for CED Higgs
production in the SM4.
The determination of spin-parity quantum numbers of Higgs bosons at the
LHC and  the advantages provided by the CED process
is briefly discussed in \refse{sec:spin}.
Finally, \refse{sec:conclusions} contains our conclusions.

%%%%%%%%%%%%%%%%%%%%%%%%%%%%%%%%%%%%%%%%%%%%%%%%%%%%%%%%%%%%%%%%%%%%%%%%%%%%%%%

\subsection*{Luminosity scenarios}

Analogous to \citeres{diffH,ismd,eds09}, 
in our numerical analysis below we 
consider four scenarios for the integrated luminosities and the
experimental conditions for CED processes at the LHC.
In order to illustrate the physics gain that could be expected if a
more optimistic scenario were to be realized (see the discussion in
Sect.~5 of \citere{diffH}), we also include in the
following a discussion of the cross sections and experimental
efficiencies with scenarios where the event rates are higher by a factor 
of~2 (due to improvements on the experimental side and possibly higher
signal rates), denoted by ``eff$\times 2$''.
We furthermore assume for sake of simplicity a center-of-mass energy of
$\sqrt{s} = 14 \tev$ at the LHC. Lower energies and correspondingly
lower cross sections would require correspondingly higher luminosities.

\newcommand{\sixoo}{60 \ifb}
\newcommand{\sixooeff}{60 \ifb\,eff$\times2$}
\newcommand{\sixooo}{600 \ifb}
\newcommand{\sixoooeff}{600 \ifb\,eff$\times2$}

\begin{itemize}
\item \underline{\sixoo :}\\
An integrated LHC luminosity of ${\cal L} = 2 \times 30~{\rm fb}^{-1}$,
corresponding roughly to three years of running at an instantaneous
luminosity
${\cal L} \sim 10^{33} \, {\rm cm}^{-2} \, {\rm s}^{-1}$ by both ATLAS
and CMS. With such a luminosity the effect of pile-up is not negligible
but can be safely kept under control. 

\bigskip
\item \underline{\sixooeff :}\\
The same integrated LHC luminosity as in the above scenario but with 
event rates that are higher by a factor of 2 (see the discussion of possible
improvements and theoretical uncertainties in \citere{diffH}). 

\item \underline{\sixooo :}\\
An integrated LHC luminosity of ${\cal L} = 2 \times 300~{\rm fb}^{-1}$
and the same efficiency factors as in the scenario with
${\cal L} = 60~{\rm fb}^{-1}$.
This corresponds roughly to three years of running at an instantaneous
luminosity $\cL \approx 10^{34}\,{\rm cm}^{-2} \, {\rm s}^{-1}$ by both
ATLAS and CMS.

\item \underline{\sixoooeff :}\\ 
The same integrated LHC luminosity as in the scenario 
with ${\cal L} = 2 \times 300~{\rm fb}^{-1}$
but with 
event rates that are higher by a factor of 2.
\end{itemize}

%%%%%%%%%%%%%%%%%%%%%%%%%%%%%%%%%%%%%%%%%%%%%%%%%%%%%%%%%%%%%%%%%%%%%%%%%%%%%%%
%%%%%%%%%%%%%%%%%%%%%%%%%%%%%%%%%%%%%%%%%%%%%%%%%%%%%%%%%%%%%%%%%%%%%%%%%%%%%%%

\section{Update of the MSSM analysis}
\label{sec:mssm}

\subsection{Tree-level structure and higher-order corrections}
\label{sec:mssmhiggs}

As it is well known, 
unlike in the SM, in the MSSM {\it two} Higgs doublets
are required.
One obtains five physical states, the neutral $\cp$-even Higgs bosons
$h, H$, the $\cp$-odd Higgs boson $A$, and the charged Higgs bosons
$H^\pm$. Furthermore there are three unphysical Goldstone boson states,
$G^0, G^\pm$.
At the tree level, the Higgs sector can be described with the help of two  
independent parameters, usually chosen as 
the ratio of the two
vacuum expectation values,  
$\tb \equiv v_2/v_1$, and $\MA$, the mass of the $\cp$-odd $A$ boson.

In the decoupling limit, which is typically reached for $\MA \gsim 150 \gev$
(depending on $\tb$), the heavy MSSM
Higgs bosons are nearly degenerate in mass, 
$\MA \approx \MH \approx \MHp$.  The
couplings of the neutral Higgs bosons to SM gauge bosons are
proportional to%
\footnote{
$\al$ diagonalizes the $\cp$-even Higgs sector at tree-level.
}%
\begin{equation}
VVh \sim \Sba\,, VVH \sim \Cba\,,
\qquad (V = Z, W^\pm)~, 
\end{equation}
while the coupling $VVA \equiv 0$ at tree level.
In the decoupling limit one finds $\be - \al \to \pi/2$, 
i.e.\ $\Sba \to 1$, $\Cba \to 0$.
Consequently, for $\MA \gsim 150 \gev$ one finds the
following decay patterns for the neutral MSSM Higgs bosons at tree-level:
\begin{itemize}
\item[$h$:]
the light $\cp$-even Higgs boson has SM-like decays. Due to its upper mass
limit of $\Mh \lsim 135$ $\gev$~\cite{mhiggsAEC} (see below), a
non-negligible decay to weak gauge bosons only occurs in a limited window
of $\Mh$ values close to this upper limit.
\item[$H,A$:]
compared to a SM Higgs boson with mass $\MHSM \gsim 150 \gev$, which 
would decay predominantly into SM gauge bosons, 
the decays of $H, A$ to SM gauge bosons are strongly suppressed. In turn, the
branching ratios to $b \bar b$ and $\tau^+\tau^-$ are much larger in
this mass range compared to the SM case. As a rule of
thumb, $\br(H,A \to b \bar b) \approx 90\%$ and 
$\br(H,A \to \tau^+\tau^-) \approx 10\%$, if SUSY particles (such as
charginos and neutralinos) are too heavy to be produced in the decays of
$H$ and $A$.
\end{itemize}

Higher-order corrections in the MSSM Higgs sector are 
in general quite large. In particular, higher-order corrections 
give rise to an upward shift of the upper bound on the light $\cp$-even 
Higgs-boson mass from the tree-level value, $\Mh \leq \MZ$, to about
$\Mh \lsim 135 \gev$~\cite{mhiggsAEC}. Besides the impact on
the masses, large higher-order corrections also affect the Higgs-boson
couplings.  For the evaluation of the theoretical predictions for the  
relevant observables in the MSSM Higgs sector we use the code
{\tt FeynHiggs}~\cite{feynhiggs,mhiggslong,mhiggsAEC,mhcMSSMlong}.

In order to fix our notation, we list the conventions for the inputs
from the scalar top and scalar bottom sector of the MSSM:
the mass matrices in the basis of the current eigenstates 
$\tilde t_L, \tilde t_R$ and
$\tilde b_L, \tilde b_R$ are given by
\begin{eqnarray}
\label{stopmassmatrix}
{\cal M}^2_{\tilde t} &=&
  \left( \begin{array}{cc}
  \MSQ^2 + \mt^2 + \cos 2\beta 
                           (\frac{1}{2} - \frac{2}{3} \sw^2) \MZ^2 &
      \mt \Xt \\
      \mt \Xt &
      \MstR^2 + \mt^2 + \frac{2}{3} \cos 2\beta \sw^2 \MZ^2 
  \end{array} \right), \\
&& \nonumber \\
\label{sbotmassmatrix}
{\cal M}^2_{\tilde b} &=&
  \left( \begin{array}{cc} \MSQ^2 + \mb^2 + \cos 2\beta 
                               (-\frac{1}{2} + \frac{1}{3} \sw^2) \MZ^2 &
      \mb \Xb \\
      \mb \Xb &
      \MsbR^2 + \mb^2 - \frac{1}{3} \cos 2\beta \sw^2 \MZ^2 
  \end{array} \right),
\end{eqnarray}
where 
\begin{equation}
\mt \Xt = \mt (\At - \mu \cot\beta) , \quad
\mb \Xb = \mb\, (\Ab - \mu \tb) .
\label{eq:mtlr}
\end{equation}
Here $\At$ denotes the trilinear Higgs--stop coupling, $\Ab$ denotes the
Higgs--sbottom coupling, and $\mu$ is the higgsino mass parameter.
As an abbreviation we will use
\BE
\msusy \equiv \MSQ = \MstR = \MsbR~.
\end{equation}

\medskip
The relation between the bottom-quark mass and the Yukawa coupling
$h_b$, which also controls the interaction between the Higgs fields and
the sbottom quarks, reads at lowest order $\mb =h_b v_1$. 
This relation is affected at \onel\ order by large radiative
corrections \cite{deltamb1,deltamb2,deltamb2b,deltamb3},
leading to a replacement of 
\begin{align}
\mb &\to \frac{\mb}{1 + \db}
\end{align}
as the dominant effect (see also
\citeres{mhiggsEP4,mhiggsEP4b,mhiggsFD2,deltab2L,dbnew}).  
The quantity $\db$ exhibits a 
parametric dependence $\db \propto \tb \, \mu$, where
the explicit form of $\db$ in the limit of $\msusy \gg \mt$ and 
$\tb \gg 1$ can be found in \citere{deltamb1}.
The CED channel, $pp \to p \oplus \phi \oplus p$ with $\phi \to b \bar b$
($\phi = h,H$) 
receives important contributions from the $\db$ corrections via the
bottom Yukawa coupling. Because of the pronounced $\mu$~dependence we
will show the results for different values of~$\mu$.

%%%%%%%%%%%%%%%%%%%%%%%%%%%%%%%%%%%%%%%%%%%%%%%%%%%%%%%%%%%%%%%%%%%%%%%%%%%%%%%

\subsection{Updates with respect to the previous analyses}
\label{sec:update}

In \citere{diffH} we have presented detailed results on signal and
background predictions of CED production of the $h$ and $H$. The results
shown here differ from the previous ones in various respects, which we
briefly summarize in the following.

%%%%%%%%%%%%%%%%%%%%%%%%%%%%%%%%%%%%%%%%%%%%%%%%%%%%%%%%%%%%%%%%%%%%%%%%%%%%%%%

\subsubsection*{Calculation of signal and \boldmath{$b\bar b$} background}

Following the studies in \citere{shuv} we use here a modified formula
(with respect to \citere{diffH}) for the overall physics background to
the $0^{++}$ Higgs signal in the $b \bar b$ mode 
\begin{align}
\frac{{\rm d}\si^B}{{\rm d} M}
\approx 0.5 \, {\rm fb/GeV} \left[
A \KL \frac{120 \gev}{M} \KR^6 
+ \frac{C}{2} \KL \frac{120 \gev}{M} \KR^8 \right] ~,
\label{eq:backbb}
\end{align}
with $A=0.92$ and $C=C_{\mathrm {NLO}}=0.48 - 0.12 \times (\ln(M/120 \gev))$.
The expression (\ref{eq:backbb}) holds for a mass window $\Delta M = 4-5\gev$
and summarizes several types of backgrounds: the prolific
$gg^{PP}\to gg$ subprocess can mimic $b\bar b$ production due
to the misidentification of the gluons as $b$ jets;
an admixture of $|J_z|=2$ production;
the radiative $gg^{PP}\to b\bar b g$ background;
due to the non-zero $b$-quark
mass there is also a contribution to the $J_z=0$
cross section of order $\mb^2/E_T^2$.
The first term in the square brackets corresponds
to the first three background sources~\cite{diffH}, evaluated
for $P_{g/b}=1.3 \%$, where $P_{g/b}$ is
the probability to misidentify a gluon as a $b$-jet for a $b$-tagging
efficiency of 60\%.%
\footnote{Further improvements in the experimental analysis
could allow to reduce $P_{g/b}$.}%
~The second term describes the background associated with bottom-mass
terms in the Born amplitude.
The NLO correction suppresses this contribution
by a factor of about 2, or more for larger masses~\cite{shuv}.
As discussed in detail in \citere{diffH}, \refeq{eq:backbb} is only
an approximative formula. A more realistic approach is to implement
all the background processes in a Monte Carlo program and to perform an
analysis at the detector level (as it was done for the signal
process). 
%However, this goes beyond the scope of our paper.}
However none of the background processes mentioned in \citere{diffH} has been 
implemented in any Monte Carlo event generator so far.
Such a dedicated background analysis goes beyond the scope of our paper.

It should be kept in mind that the main experimental challenge of
running at high luminosity,\ 
$\cL \sim 10^{34} \, {\rm cm}^{-2} \, {\rm s}^{-1}$,
is the effect of pile-up, which can generate fake signal
events within the
acceptances of the proton detectors
as a result of the coincidence of two or more separate
interactions in the same bunch crossing,
see \citeres{CMS-Totem,FP420,diffH,CLP} for details.
As established in \citeres{CLP,kp2,kp3} we can expect that
this overlap background can be brought
under control by using dedicated fast-timing proton detectors with
a few pico-second resolution (see \citere{FP420})
and  additional experimental cuts.
Note also that in the analysis strategy of \citere{diffH}
the event selections and cuts were
imposed such as to maximally reduce the pile-up background.
Similarly to \citere{diffH}, 
based on the anticipated improvements for a reduction of the
overlap backgrounds down to a tolerable level,
here the pile-up effects are assumed to be negligible after applying all
the cuts suppressing the pile-up. Consequently, the remaining
pile-up backgrounds are not included in our numerical studies performed
in the present paper.

{}From a perspective of an even further future, 
there is an idea to double the
Level-1 trigger latency in CMS which would allow the detectors at 420~m
to be included in the Level-1 trigger decision. Although no dedicated
study of the efficiency of such a trigger or the effect of pile-up on it
has so far been made, we believe that a fair increase of the
statistical significance could be expected in this case.

The signal cross section is calculated on the basis of the 
prediction for the production of a SM Higgs boson in CED together with an
appropriate rescaling using the partial widths of the neutral $\cp$-even
Higgs bosons of the MSSM into gluons, $\Ga(\phi \to gg)$ ($\phi = h,H$),
as implemented in {\tt FeynHiggs}
(details are given in \citere{diffH}).
The predictions within {\tt FeynHiggs} have been updated 
(from the version {\tt 2.3.0-beta} used in \citere{diffH} to the version 
{\tt 2.7.1} used here) by  
(see the ``change log'' in \citere{fh-web} for details)
\begin{itemize}

\item a change in the bottom Yukawa coupling employed for the bottom
  loop contribution. The running bottom mass is now evaluated at the
  scale of the bottom mass, $\mb(\mb)$. This scale choice was found
  to yield smaller higher-order corrections compared to the previous 
  parametrisation in terms of $\mb(\mt)$. 
  This change leads to a sizable enhanced
  $gg \to h (H)$ production rate at lower (all) $\MA$,
\item the inclusion of an improved version of the $\db$
  corrections~\cite{dbnew} to the bottom loop in the $\phi \to gg$
  calculation:
more higher-order contributions to $\db$ are taken into account, leading
  in our case to a further
  effective enhancement of the bottom Yukawa coupling, 
  and thus to an enhanced $gg \to h (H)$ production rate at lower (all) $\MA$,
\item the change to a running top mass, effectively decreasing the
  top-loop contribution, or, conversely, increasing the relative
  bottom-loop contribution, and effectively to a slightly
  enhanced $gg \to h (H)$ production rate at lower (all) $\MA$,
\item further amendments having a smaller numerical impact.
\end{itemize}
The predictions have been improved in particular
in the region of low $\MA$ and/or large $\mu$ and large $\tb$.

Since the publication of \citere{diffH} there were some further
developments in the calculations of the CED cross-sections
concerning both, the hard matrix element (see \citeres{mrw,cf,acf})
and the so-called soft absorptive corrections and soft-hard
factorization breaking effects (see \citeres{kmrf,nns2,nnns} for details
and references).
However, at the present stage
we do not see sufficient reason to revise
the result used in \citere{diffH} for the calculation
of the effective exclusive $gg$-luminosity, which determines the
rates of signal and background events, see also \citere{kmrf}.
It is worth emphasizing in this context that the effective luminosity
cancels in the signal-to-background ratio for the non-pile-up
background. Thus, although the overall uncertainty in the calculation
of the CED cross section could be as large as a factor of $\sim 2.5$,
the effect of this uncertainty on the Higgs  discovery contours is
comparatively weak.
It should furthermore be noted 
that there is a good chance that the accuracy of the predictions
will improve after the early runs at the LHC which will allow
a detailed test of the theoretical formalism, see \citere{early}.

\begin{figure}[htb!]
%\vspace{0.5em}
\begin{center}
\includegraphics[width=14cm,height=8.8cm]
                {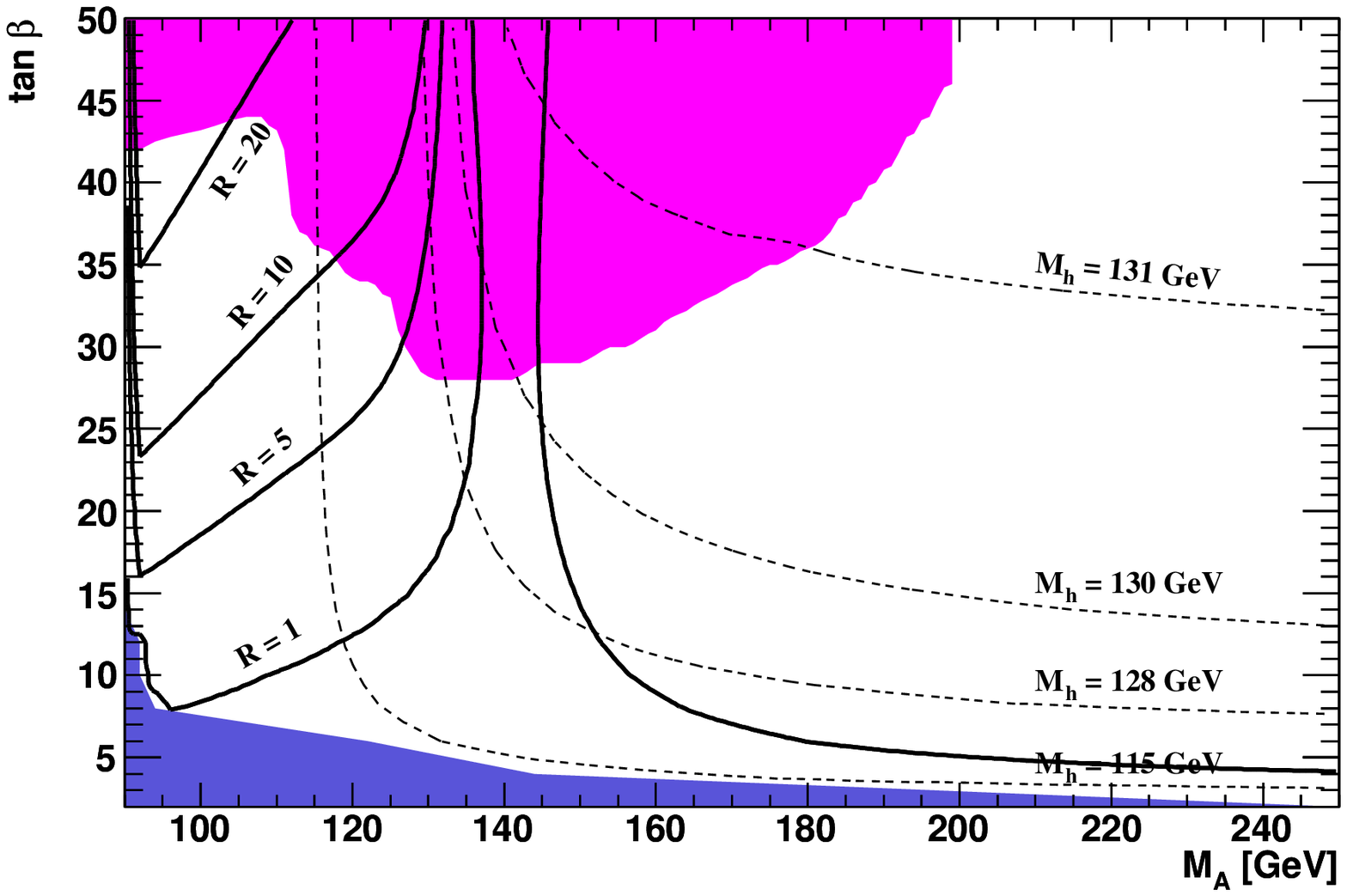}
\includegraphics[width=14cm,height=8.8cm]
                {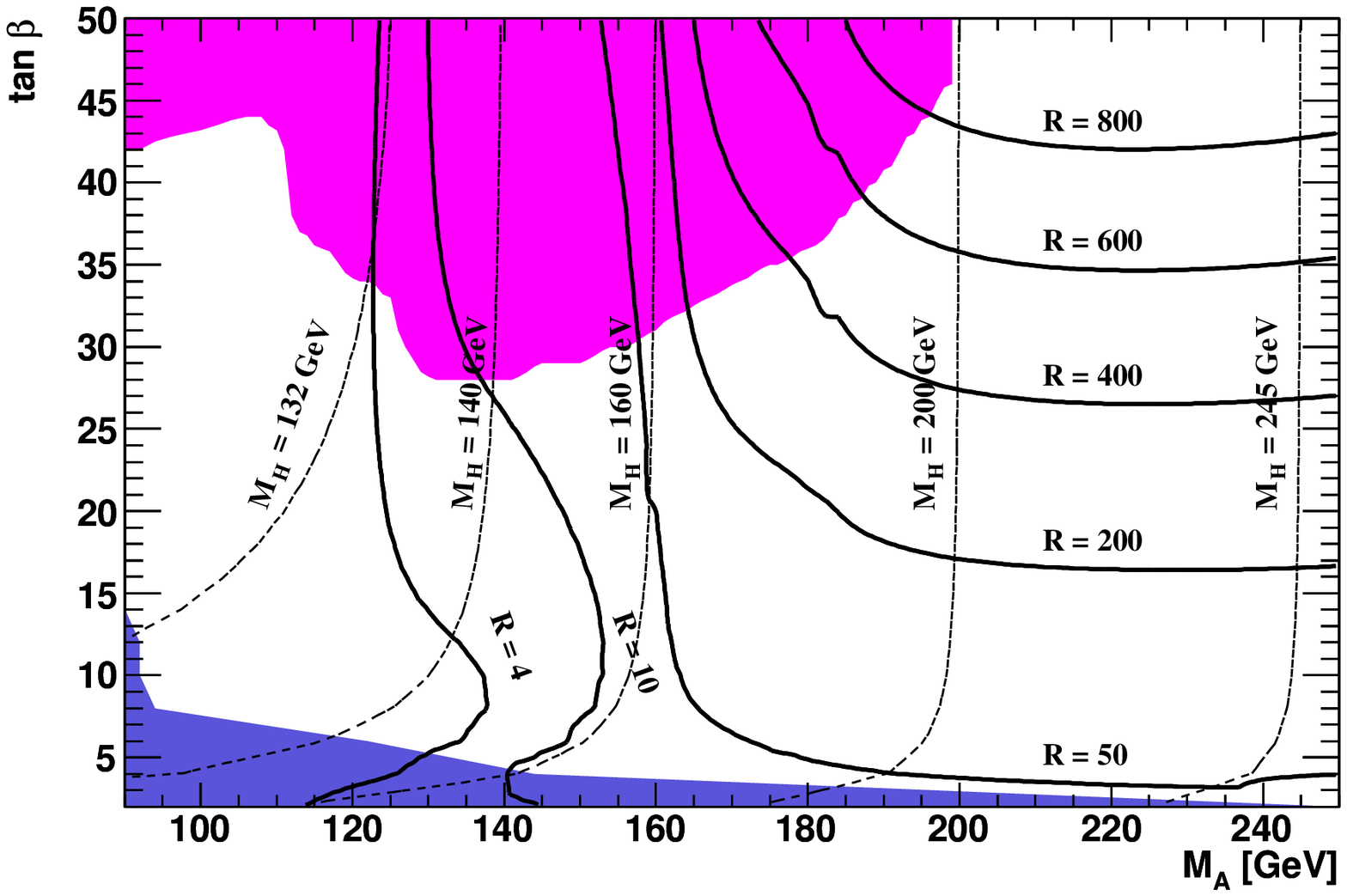}
%\vspace{1em}
\caption{
Contours for the ratio of signal events in the MSSM to those in the SM
in the $h(H) \to b \bar b$ channel in CED production in the $\MA$--$\tb$
plane are shown in the upper (lower) plot. The ratio is displayed in the
$\Mhmax$ benchmark scenario with $\mu = +200 \gev$. 
The values of the mass of the light (heavy) $\cp$-even Higgs boson, 
$\Mh$ ($\MH$), are indicated by contour lines. The dark shaded (blue)
region corresponds to the parameter region that
is excluded by the LEP Higgs 
searches, the lighter shaded (pink) areas are excluded by Tevatron Higgs
searches. 
}
\label{fig:ratios}
\end{center}
\end{figure}

To summarize the effect of the various changes in the 
evaluation of the MSSM contributions to the scaling factors that 
we apply to the production rate of a SM Higgs boson in CED
background calculations we
show in the upper (lower) plot of \reffi{fig:ratios} the contours
for the ratio of signal 
events in the MSSM to those in the SM in the $h(H) \to b \bar b$ channel
in CED production in the $\MA$--$\tb$ plane. The parameters are fixed
according to the $\Mhmax$ benchmark scenario~\cite{benchmark2} with 
$\mu = +200 \gev$ (see \refeq{mhmax} below). 
These plots are updated from Fig.~2 (upper plot) and
Fig.~7 (upper plot) of \citere{diffH}.

%%%%%%%%%%%%%%%%%%%%%%%%%%%%%%%% Begin FIGURE %%%%%%%%%%%%%%%%%%%%%%%%%%%%%%%%%
%%%%%%%%%%%%%%%%%%%%%%%%%%%%%%%% End FIGURE %%%%%%%%%%%%%%%%%%%%%%%%%%%%%%%%%%%

Comparing our updated results with the plots in \citere{diffH} 
one can see that the effect of the improved $gg \to h/H$ calculation 
gives rise to an increase of the MSSM CED Higgs production channel 
everywhere, while the size of the enhancement factor depends on 
the parameter region and the
channel. In the case of the $h$ production and decay into $b \bar b$, 
we find an enhancement compared to our previous analysis of up to a factor 
of 2 for $\MA < 140 \gev$, while for $\MA > 140 \gev$ the change does not
exceed the level of 10\%. For $H$ production and decay into $b \bar b$, 
we find an increase up to the level of 50\% for $\MA < 130\gev$, while 
an enhancement of up to a factor 3 is possible for larger $\MA$.
As we will discuss below, the enhanced production rates obtained from
the improved theoretical predictions for $gg \to h/H$ give rise to an 
enlargement of the parameter regions that can be covered with CED
Higgs production processes within the MSSM at the $5\,\si$ and the
$3\,\si$ level. On the other hand, the parameter regions excluded by the
Higgs searches at the Tevatron have also increased as compared to our
previous analysis. We show the contours of $5\,\si$ and $3\,\si$
significances for the central values of the signal and background
predictions discussed above, i.e., the remaining theoretical uncertainty in
the signal-to-background ratios is not explicitly taken into account.

%%%%%%%%%%%%%%%%%%%%%%%%%%%%%%%%%%%%%%%%%%%%%%%%%%%%%%%%%%%%%%%%%%%%%%%%%%%%%%%

\subsubsection*{Benchmark scenarios}

Due to the large number of MSSM parameters, a number of benchmark 
scenarios~\cite{benchmark2,benchmark3}
have been used for the interpretation of MSSM Higgs
boson searches at LEP~\cite{LEPHiggsSM,LEPHiggsMSSM} and at the
Tevatron~\cite{TevHiggsSM,TevHiggsMSSM}.
Two of the most commonly used benchmark scenarios for the
$\cp$-conserving MSSM~\cite{benchmark2,benchmark3} were used in
\citere{diffH}: 

\begin{itemize}

\item
\ul{The $\Mhmax$ scenario:}\\[.5em]
In this scenario the parameters are chosen such that the mass of the
light $\cp$-even Higgs boson acquires its maximum possible values 
as a function of $\tb$ (for fixed $M_{\rm SUSY}$, $\mt$,
and $\MA$ set to its maximum value, $\MA = 1$~TeV).
This was used in particular to obtain conservative $\tb$ 
exclusion bounds~\cite{tbexcl} at LEP~\cite{LEPHiggsMSSM}.
The parameters are:
\begin{eqnarray}
&& \mt = 173.3 {\rm ~GeV}, \quad \msusy = 1 {\rm ~TeV}, \quad
\mu = 200 {\rm ~GeV}, \quad M_2 = 200 {\rm ~GeV}, \nonumber \\
\label{mhmax}
&& \Xt = 2\, \msusy  \quad
   \Ab = \At, \quad m_{\tilde g} = 0.8\,\msusy~.
\end{eqnarray}

\item
\ul{The no-mixing scenario:}\\[.5em]
This benchmark scenario is the same as the $\Mhmax$ scenario, but with
vanishing mixing in the $\tilde t$~sector and with a higher SUSY mass
scale (the latter has been chosen to avoid conflict with the exclusion
bounds from the LEP Higgs searches~\cite{LEPHiggsMSSM,LEPHiggsSM}),
\begin{eqnarray}
&& \mt = 173.3 {\rm ~GeV}, \quad \msusy = 2 {\rm ~TeV}, \quad
\mu = 200 {\rm ~GeV}, \quad M_2 = 200 {\rm ~GeV}, \nonumber \\
\label{nomix}
&& \Xt = 0  \; \quad
   \Ab = \At, \quad m_{\tilde g} = 0.8\,\msusy~.
\end{eqnarray}

\end{itemize}
As discussed above, in order to study the impact of potentially large
corrections in the $b/\tilde{b}$ sector it is useful to vary the absolute 
value and sign of the parameter $\mu$. 
The other MSSM parameters that are not specified above have only a minor
impact on MSSM Higgs-boson phenomenology. In our numerical analysis
below we fix them such that all soft SUSY-breaking parameters in the
diagonal entries of the sfermion mass matrices are set to $\msusy$, and
the trilinear couplings for all sfermions are set to $\At$.

\smallskip
The benchmark scenarios defined above had been designed to facilitate
the Higgs-boson searches at LEP, the Tevatron and the LHC.
Consequently, they do not pay particular attention to the compliance
with other constraints, such as electroweak precision observables,
$B$~physics observables and the abundance of cold dark matter (CDM),
see, for instance, \citere{EWPOBPOCDM} and references therein.
In \citere{CDM} four new MSSM benchmark scenarios based on the
NUHM2~\cite{nuhm2} were defined%
\footnote{
The NUHM2, which stands for ``Non-Universal Higgs Model type 2'',
assumes common mass scales for scalars, $m_0$, for
fermions, $m_{1/2}$, and for trilinear couplings, $A_0$, 
at the Grand Unification scale,  $M_{\rm GUT} \sim 2 \times 10^{16} \gev$. 
However, the two scalar soft SUSY-breaking parameters of the Higgs
sector are assumed to be unconnected to $m_0$. Together with $\tb$ the
NUHM2 is described by six free parameters (and the sign of~$\mu$).}%
. In these scenarios
the abundance of the lightest SUSY particle, the lightest neutralino, in
the early universe is compatible within the $\MA$--$\tb$ plane
with the CDM constraints as measured by WMAP~\cite{WMAP}.
The parameters chosen for the benchmark planes are also in
agreement with electroweak precision and $B$-physics constraints, see
\citere{CDM} for further details.
Below we will present the prospects for CED Higgs production also in
two of these four ``CDM benchmark scenarios'', labeled \pdrei\ and
\pvier. More details about their definition can be found in
\citere{CDM}.

%%%%%%%%%%%%%%%%%%%%%%%%%%%%%%%%%%%%%%%%%%%%%%%%%%%%%%%%%%%%%%%%%%%%%%%%%%%%%%%

\subsubsection*{Bounds from Higgs searches at LEP and the Tevatron}

The Higgs bosons of the SM and the MSSM have been searched for at
LEP~\cite{LEPHiggsSM,LEPHiggsMSSM} and the
Tevatron~\cite{TevHiggsSM,TevHiggsMSSM}. 
The results presented in \citere{diffH} showed the parts of the
$\MA$--$\tb$ planes that are excluded by the LEP Higgs-boson
searches. Since then the Tevatron has considerably improved its reach
for the SUSY Higgs bosons, especially for relatively low $\MA$ and large
$\tb$. 
The bounds obtained at the Tevatron (together with the previously known
LEP bounds) have been implemented into the Fortran code 
{\tt HiggsBounds}~\cite{higgsbounds} (linked to 
{\tt FeynHiggs}~\cite{feynhiggs,mhiggslong,mhiggsAEC,mhcMSSMlong} 
to provide the relevant Higgs masses and couplings).
For any parameter point provided to {\tt HiggsBounds} the code
determines whether it is
excluded at the 95\%~C.L.\ based on the published exclusion data.
These excluded regions from LEP {\em and} the Tevatron are marked in the
MSSM ($\Mhmax$, no-mixing and CDM benchmark scenarios)
and the SM4 plots shown below. 
We have used the version {\tt HiggsBounds\,1.2.0} for our evaluations.

%%%%%%%%%%%%%%%%%%%%%%%%%%%%%%%%%%%%%%%%%%%%%%%%%%%%%%%%%%%%%%%%%%%%%%%%%%%%%%%

\subsection{Updated discovery reach for neutral $\cp$-even Higgs bosons 
in the \boldmath{$\Mhmax$} and no-mixing scenarios}
\label{sec:bench}

In this section we present the updated prospects for observing the neutral
$\cp$-even MSSM Higgs bosons in CED production. We display our results
in the $\MA$--$\tb$ planes for the $\Mhmax$ and no-mixing benchmark
scenarios (the new results for the ``CDM benchmark scenarios'' can be
found in the next subsection).
As explained there, also shown in the plots 
are the parameter regions excluded by the LEP Higgs searches (as dark
shaded (blue) areas) and Tevatron Higgs-boson searches (as lighter
shaded (pink) areas) as obtained with {\tt HiggsBounds}~\cite{higgsbounds}.
Concerning the Tevatron exclusion bounds it should be noted that they
largely rely on the searches in the channel 
$b \bar b \to h,H,A \to \tau^+\tau^-$. The SM cross section
used~\cite{bbhatnnlo} for the normalization within {\tt HiggsBounds}
is evaluated using
the MRST2002 NNLO PDFs~\cite{mrst2002}. Using the updated version
MSTW2008~\cite{mstw2008} results in a reduction of the cross section
by $\sim 20\%$, which translates into weaker bounds on $\tb$ by about
$10\%$.

For each point in the parameter space we have evaluated the relevant Higgs
production cross section times
the Higgs branching ratio corresponding to the decay mode under
investigation. The Higgs-boson masses, the decay branching ratios and
the effective couplings for the production cross sections
have been calculated with the program
\fh~\cite{feynhiggs,fh-web,mhiggslong,mhiggsAEC,mhcMSSMlong}.
The resulting theoretical cross section has been multiplied by the
experimental efficiencies taking into account detector acceptances,
experimental cuts and triggers as discussed in \citere{diffH}.
The backgrounds have been estimated according to the update given in the
previous subsection.

This procedure has been carried out for four different assumptions on the
luminosity scenario, see \refse{sec:intro}, for which the $5\,\si$
discovery contours and contours for $3\,\si$ significances (see below)
have been obtained.

\medskip
We start our analysis with the production of the lighter $\cp$-even Higgs
boson, $h$, and its decay into bottom quarks. As explained in
\refse{sec:mssmhiggs}, the $hb\bar b$ coupling can be significantly
enhanced compared to the SM case in the region of relatively small $\MA$
and large $\tb$ (while in the decoupling region, $\MA \gg \MZ$, the
lighter $\cp$-even Higgs boson of the MSSM behaves like the SM Higgs
boson). 
Therefore,
in parameter regions where the relevant couplings are enhanced 
compared to the SM case, 
CED production of the lighter $\cp$-even Higgs boson of the MSSM with
subsequent decay to $b \bar b$ yields a higher event rate.

\begin{figure}[htb!]
%\vspace{0.5em}
\begin{center}
\includegraphics[width=14cm,height=8.8cm]
                {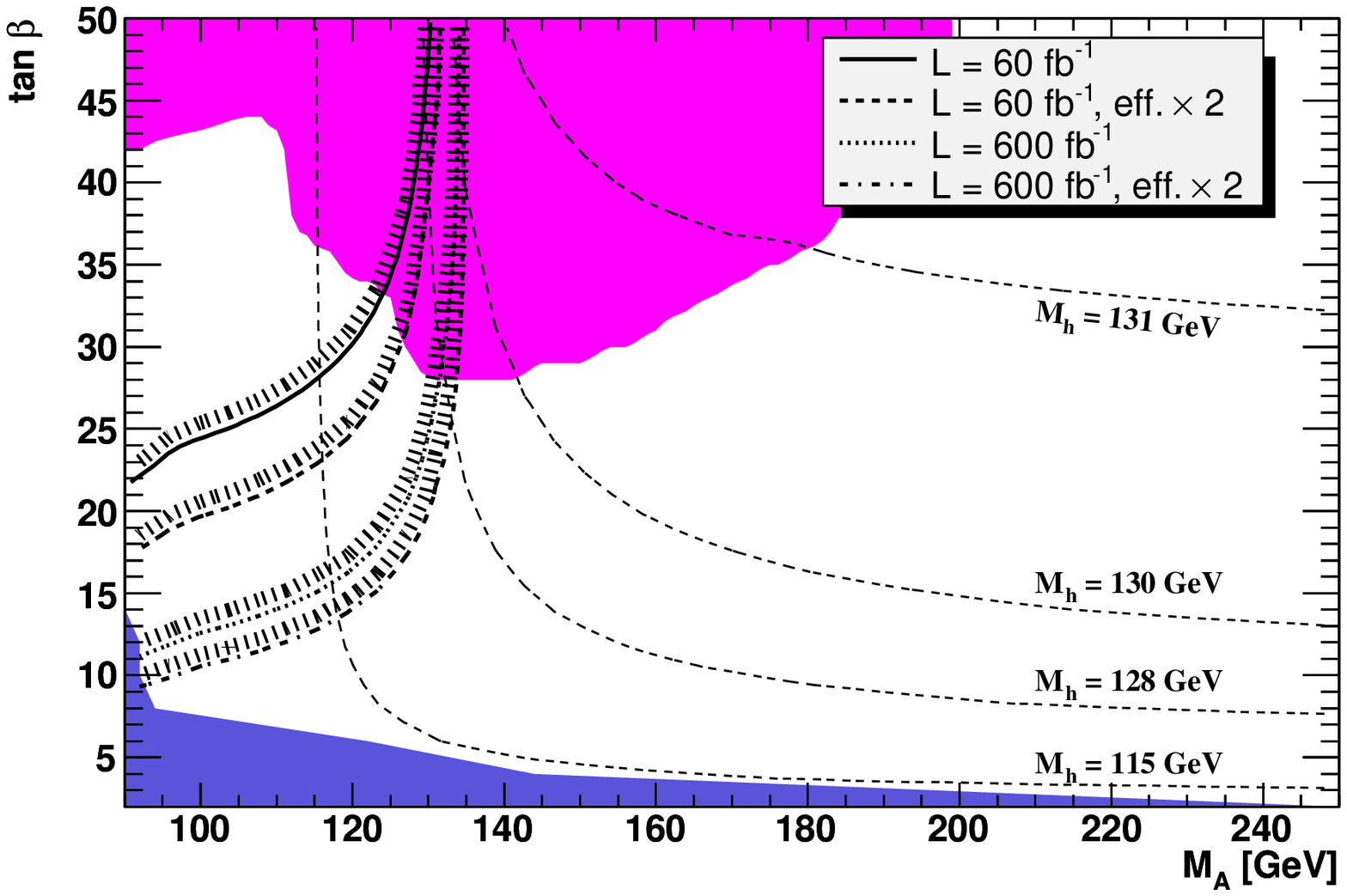}
\includegraphics[width=14cm,height=8.8cm]
                {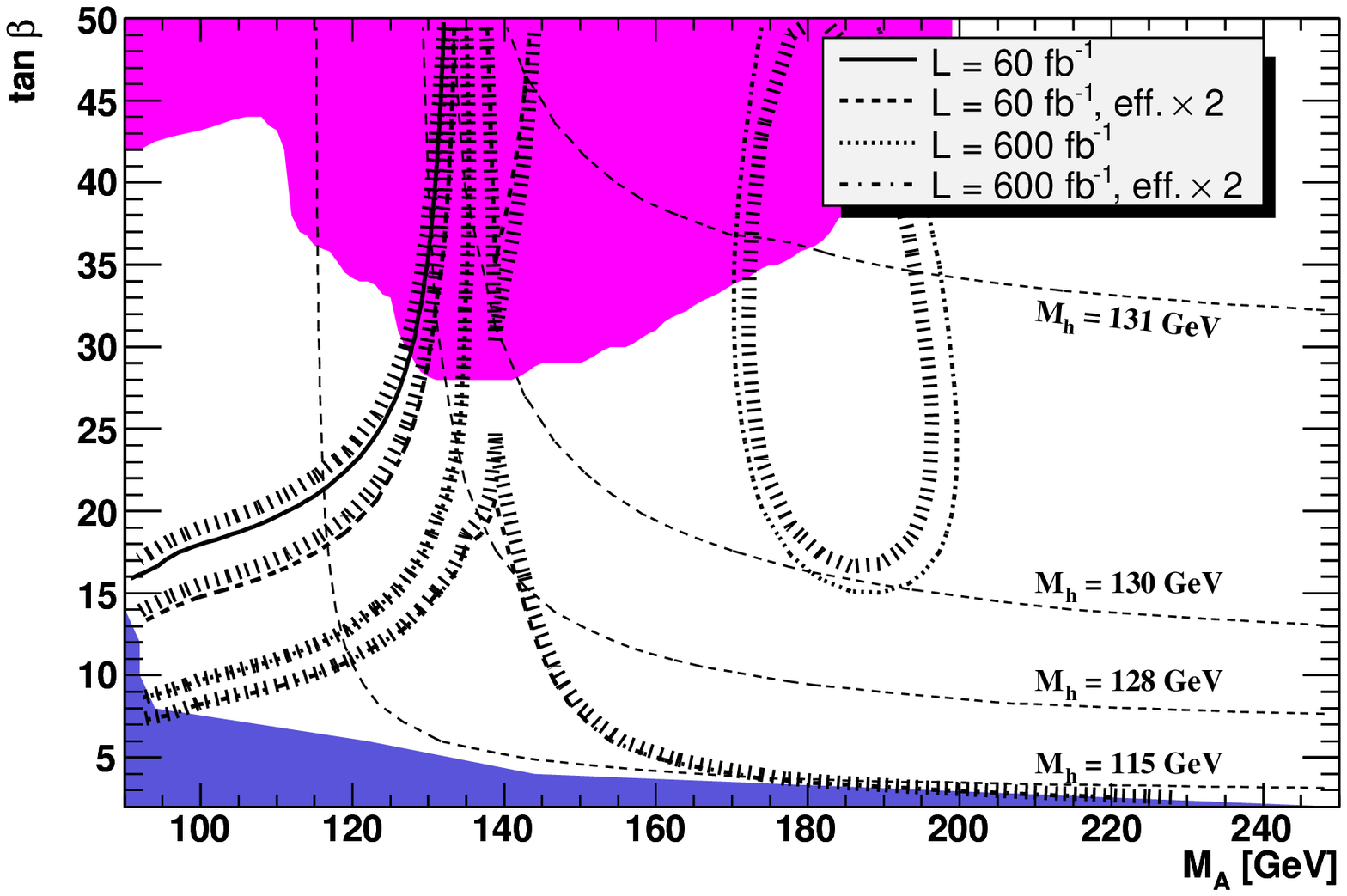}
%\vspace{1em}
\caption{
$5\,\si$ discovery contours (upper plot) and contours of $3\,\si$
statistical significance (lower plot) for the $h \to b \bar b$ channel in
CED production in the $\MA$--$\tb$ plane of the MSSM within the $\Mhmax$
benchmark scenario. The results are shown for assumed effective
luminosities (see text, combining ATLAS and CMS) of \sixoo,
\sixooeff, \sixooo\ and \sixoooeff. 
The values of the mass of the light $\cp$-even Higgs boson, $\Mh$, are
indicated by contour lines. The dark shaded (blue)
region corresponds to the parameter region that
is excluded by the LEP Higgs 
searches, the lighter shaded (pink) areas are excluded by Tevatron Higgs
searches. 
}
\label{fig:hbb1}
\end{center}
\end{figure}

In \reffi{fig:hbb1} we show the 
$5\,\si$ discovery contours (upper plot) and contours of $3\,\si$
statistical significance (lower plot) for the $h \to b \bar b$ channel in
CED production in the $\MA$--$\tb$ plane of the MSSM within the $\Mhmax$
benchmark scenario. The results are shown for assumed effective 
integrated luminosities (see text, combining ATLAS and CMS) of \sixoo,
\sixooeff, \sixooo\ and \sixoooeff. 
Since the lighter $\cp$-even Higgs boson of the MSSM is likely to be 
detectable also in ``conventional'' Higgs search channels at the LHC
(see for example \citeres{atlastdr,CMS-TDR}), a $3\,\si$ statistical
significance for the CED channel could be considered as sufficient.
The values of the mass of the light $\cp$-even Higgs boson, $\Mh$, are
indicated by contour lines. The dark shaded (blue) region corresponds to
the parameter region that is excluded by the LEP Higgs  
searches, the lighter shaded (pink) areas are excluded by Tevatron Higgs
searches. 

The $5\,\si$ discovery regions cover the parameter space of 
$\MA \le 135 \gev$ and $\tb \gsim 10$, depending on the luminosity
assumption. Some parts at larger $\tb$ are excluded by the
Tevatron Higgs-boson 
searches. The $3\,\si$ regions are much larger and cover nearly the
whole $\MA$--$\tb$ plane in the high-luminosity scenarios. The only
exception is a funnel around $\MA \approx 140 \gev$
for $\tb < 25$ or $\tb > 30$, 
where $\tb \gsim 28$ in the uncovered region is
excluded by the Tevatron Higgs searches. 
The fact that coverage up to highest $\MA$ values displayed in 
\reffi{fig:hbb1} is obtained in the high-luminosity scenarios indicates
that in this case a $3\,\si$ significance would be obtained for a SM
Higgs with a mass value corresponding to the light $\cp$-even Higgs mass
of the MSSM.
In comparison with the results presented in \citere{diffH} one can see
that the various updates of the analysis
enlarge the regions allowing a $5\,\si$ discovery as well as the ones
yielding a $3\,\si$ significance.
Similarly to the case of the ratios discussed in \refse{sec:update}, the
enhancements with respect to the previous results of \citere{diffH} are
of up to a factor of 2 for $\MA \lsim 140 \gev$, while  
they do not exceed 10\% for $\MA \gsim 140 \gev$.

%%%%%%%%%%%%%%%%%%%%%%%%%%%%%%%% Begin FIGURE %%%%%%%%%%%%%%%%%%%%%%%%%%%%%%%%%
%%%%%%%%%%%%%%%%%%%%%%%%%%%%%%%% End FIGURE %%%%%%%%%%%%%%%%%%%%%%%%%%%%%%%%%%%

\begin{figure}[htb!]
%\vspace{0.5em}
\begin{center}
\includegraphics[width=14cm,height=8.8cm]
                {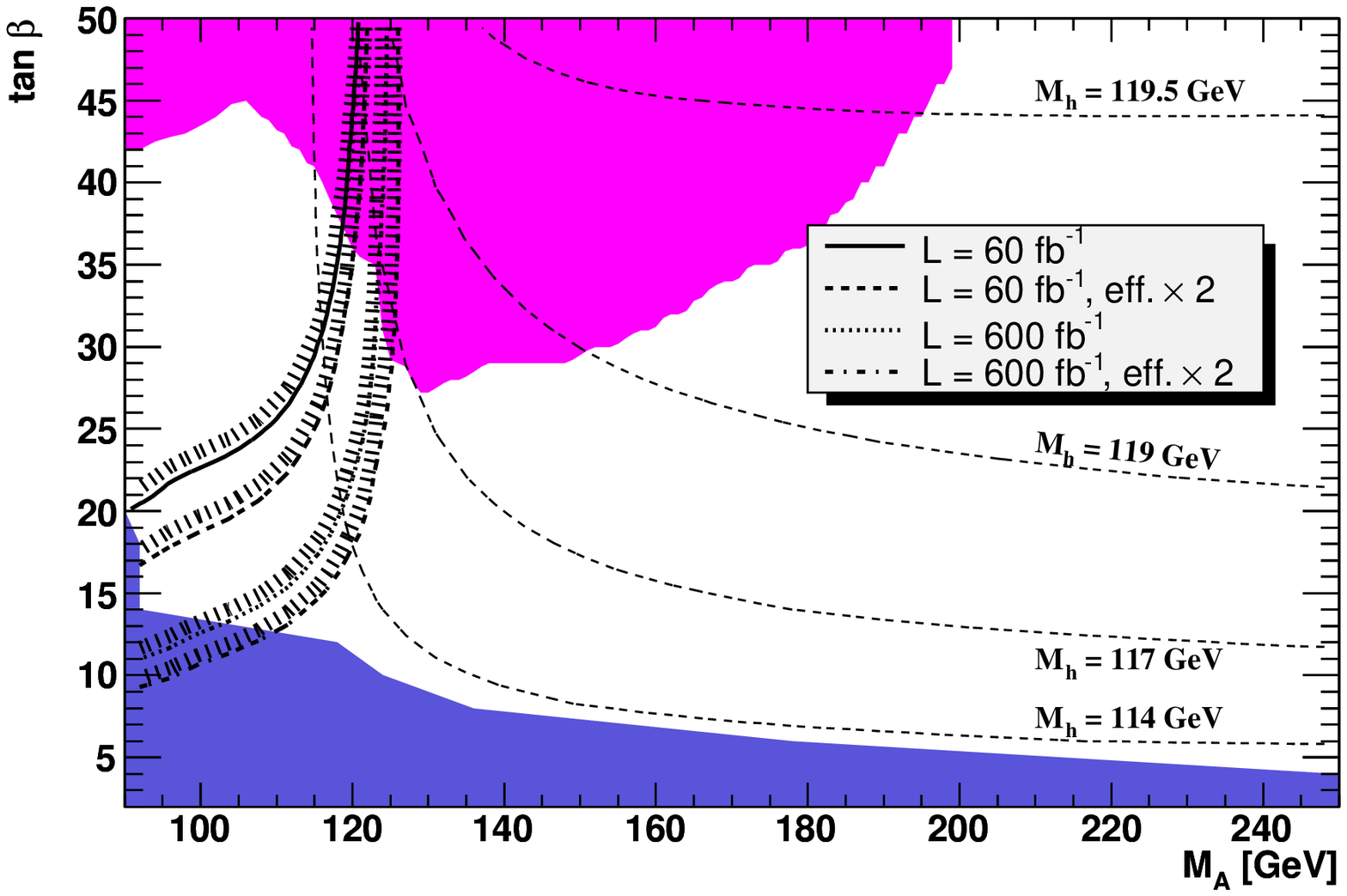}
\includegraphics[width=14cm,height=8.8cm]
                {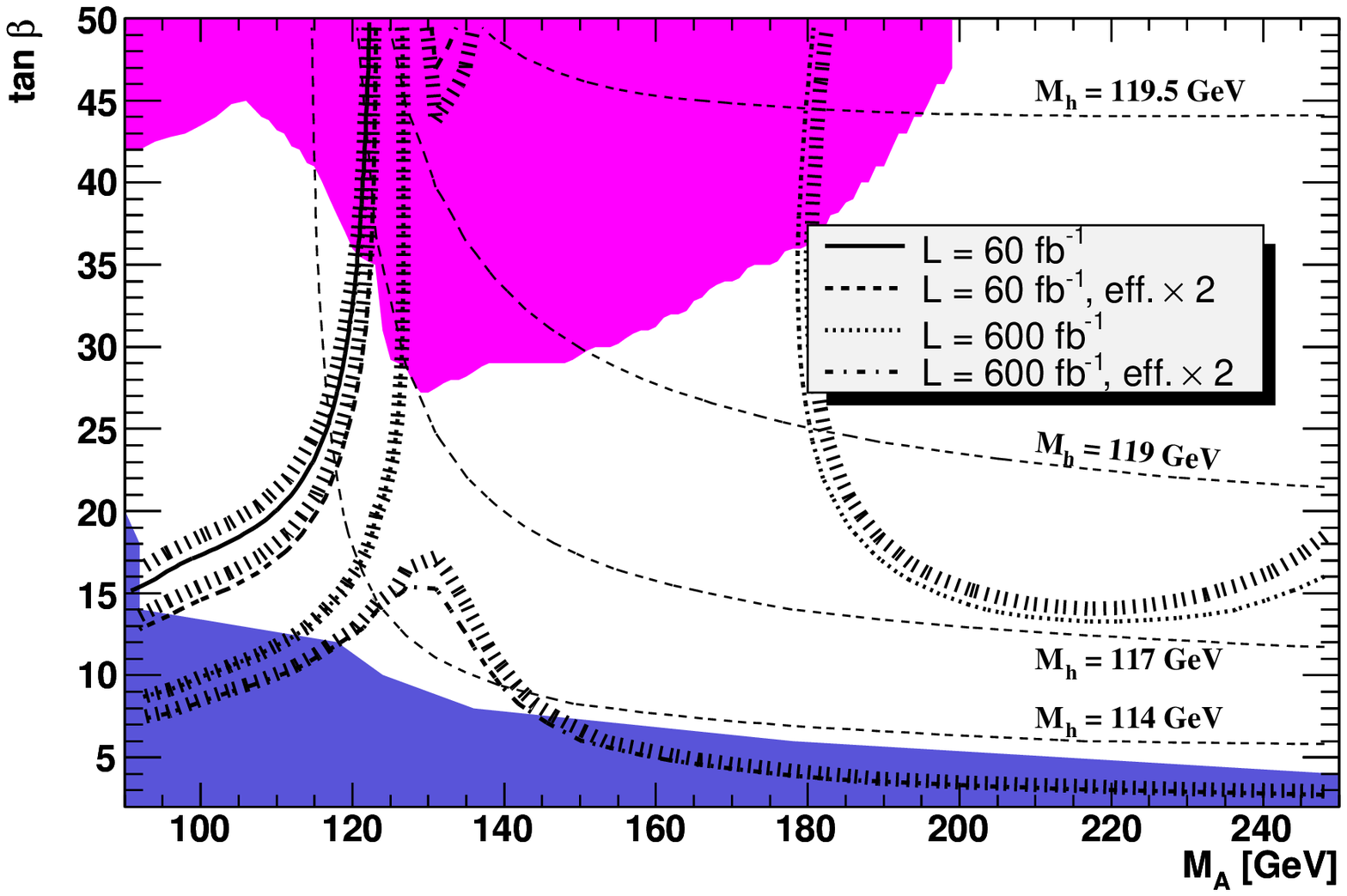}
%\vspace{1em}
\caption{
$5\,\si$ discovery contours (upper plot) and contours of $3\,\si$
statistical significance (lower plot) for the $h \to b \bar b$ channel in
CED production in the $\MA$--$\tb$ plane of the MSSM within the no-mixing
benchmark scenario. The results are shown for assumed effective
luminosities (see text, combining ATLAS and CMS) of \sixoo,
\sixooeff, \sixooo\ and \sixoooeff. 
The values of the mass of the light $\cp$-even Higgs boson, $\Mh$, are
indicated by contour lines. The dark shaded (blue)
region corresponds to the parameter region that
is excluded by the LEP Higgs 
searches, the lighter shaded (pink) areas are excluded by Tevatron Higgs
searches. 
}
\label{fig:hbb2}
\end{center}
\end{figure}

The results look very similar in the no-mixing scenario as shown in
\reffi{fig:hbb2}. The $5\,\si$ discovery regions, given in the upper
plot, end at somewhat lower $\MA$ values, $\MA \lsim 125 \gev$. 
The uncovered funnel for the $3\,\si$ significance, given in the lower
plot
has shrunk such that only $\tb \lsim 15$ and $\tb \gsim 45$ remain
uncovered. 
Concerning the experimental bounds, the LEP bounds extend to
higher $\tb$ values, reflecting the fact that much lower
$\Mh$ values are realized in this scenario.%
\footnote{
The fact that the LEP bounds do not reach 
$\Mh = 114 \gev$~\cite{LEPHiggsSM,LEPHiggsMSSM} reflects the
theory uncertainty of $\sim 3 \gev$ in $\Mh$~\cite{mhiggsAEC} 
taken into account for the LEP bounds.
}%
~The Tevatron bounds, on the other
hand, remain nearly unchanged.
The comparison with the results in \citere{diffH} 
leads to the same conclusions as in the $\Mhmax$ scenario:
the various updates enlarge the regions delimited by the $5\,\si$
  discovery contours as well as by the $3\,\si$ significance contours
(with the details discussed in the case of the $\Mhmax$ scenario).

%%%%%%%%%%%%%%%%%%%%%%%%%%%%%%%% Begin FIGURE %%%%%%%%%%%%%%%%%%%%%%%%%%%%%%%%%
%%%%%%%%%%%%%%%%%%%%%%%%%%%%%%%% End FIGURE %%%%%%%%%%%%%%%%%%%%%%%%%%%%%%%%%%%

The CED production process with subsequent decay into
bottom quarks is of particular relevance since this channel may be the
only possibility for directly accessing the $hb \bar b$ coupling,%
\footnote{
Another interesting idea to access the $b \bar b$ coupling to the
Higgs boson 
is the production via Higgs-strahlung, $V^* \to VH$ ($V = W^\pm, Z$)
in a strongly boosted system~\cite{atlastdr,bbH-boost}.
}%
~although the decay into bottom quarks is by far 
the dominant decay mode
of the lighter $\cp$-even Higgs boson in nearly the whole
parameter space of the MSSM (and it is also the dominant decay of a light 
SM-like Higgs). 
For this reason information on the bottom Yukawa coupling is important for
determining {\em any\/} Higgs-boson coupling at the LHC (rather than
just ratios of couplings), see \citeres{HcoupLHCSM,HcoupLHC120,lhc2fc}.
\reffis{fig:hbb1},~\ref{fig:hbb2} show that at the $3\,\si$ level a
significantly larger part of the $\MA$--$\tb$ plane can be covered
compared to the $5\,\si$ discovery contours. In particular, in the
``\sixoooeff'' scenario the coverage in both benchmark
scenarios extends to large $\MA$ values and small values of $\tb$.
With the exception of a small parameter region around 
$\MA \lsim 140 (130) \gev$ and low $\tb$ values, in the $\Mhmax$ 
(no-mixing) scenario the
whole $\MA$--$\tb$ plane of the MSSM (and also the case of a light
SM-like Higgs) can be covered with the CED process in this
case. This important result implies that if the CED channel can be utilized 
at high instantaneous
luminosity (which requires in particular that pile-up background
is brought under control, see the discussion in
\refse{sec:update} and in \citere{diffH}) there is a good chance to
detect the  
lighter $\cp$-even Higgs boson of the MSSM in this channel with
subsequent decay into bottom quarks, yielding crucial information on the
properties of the new state.

\begin{figure}[htb!]
%\vspace{0.5em}
\begin{center}
\includegraphics[width=14cm,height=8.5cm]
                {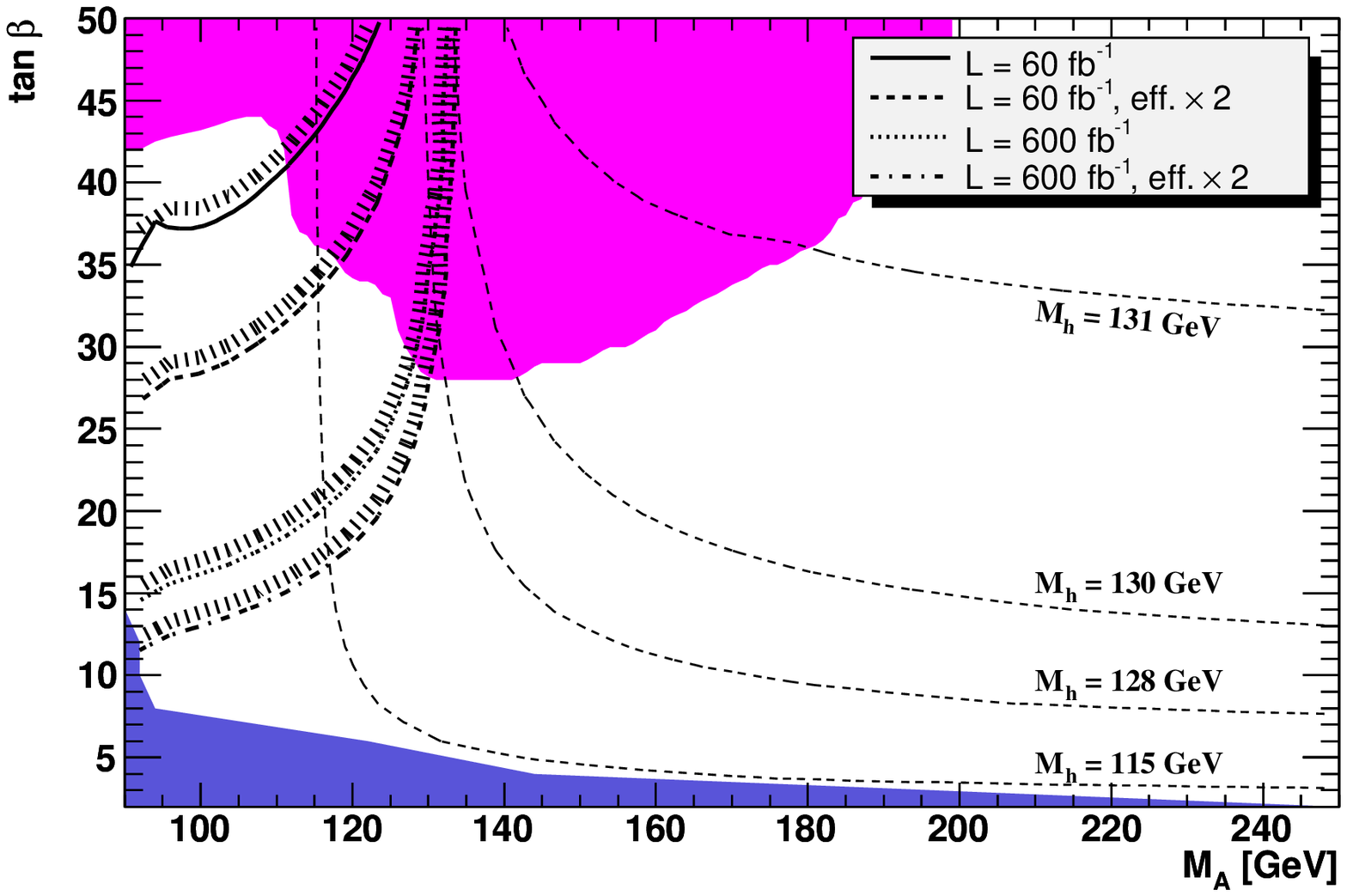}
\includegraphics[width=14cm,height=8.5cm]
                {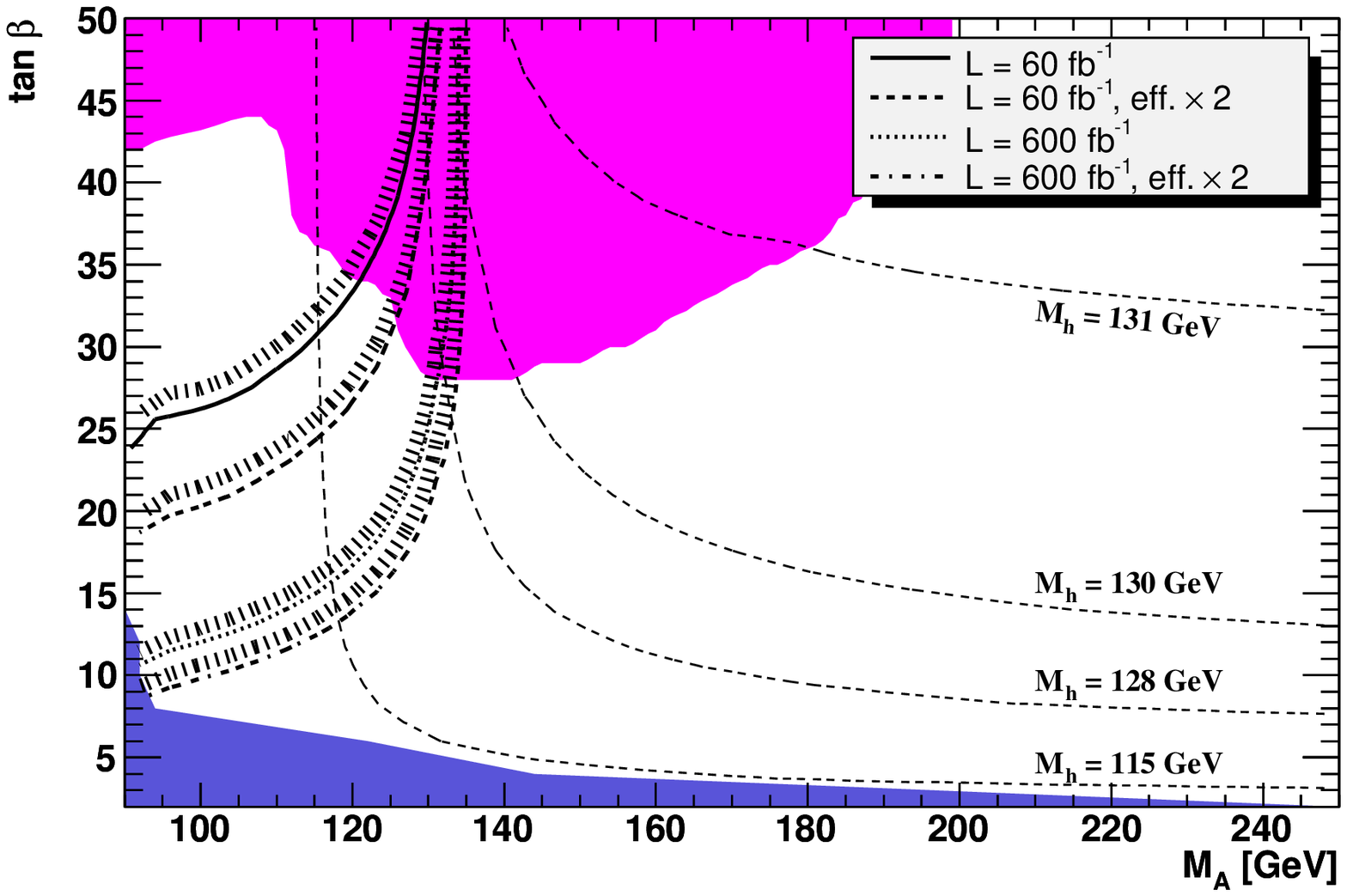}
%\vspace{1em}
\caption{
$5\,\si$ discovery contours (upper plot) and contours of $3\,\si$
statistical significance (lower plot) for the $h \to \tau^+\tau^-$ channel in
CED production in the $\MA$--$\tb$ plane of the MSSM within the $\Mhmax$
benchmark scenario. The results are shown for assumed effective
luminosities (see text, combining ATLAS and CMS) of \sixoo,
\sixooeff, \sixooo\ and \sixoooeff. 
The values of the mass of the light $\cp$-even Higgs boson, $\Mh$, are
indicated by contour lines. The dark shaded (blue)
region corresponds to the parameter region that
is excluded by the LEP Higgs 
searches, the lighter shaded (pink) areas are excluded by Tevatron Higgs
searches. 
}
\label{fig:htautau}
\end{center}
\end{figure}

The updated prospects for the $h \to \tau^+\tau^-$ channel in the
$\Mhmax$ scenario are presented in \reffi{fig:htautau}. The $5\,\si$
contours, shown in the upper plot, cover slightly smaller regions than
the corresponding contours for the $h \to b \bar b$ channel in 
\reffi{fig:hbb1}. 
The resulting features are similar for the $3\,\si$ contours shown in
the lower plot, with the exception of the high luminosity scenarios, 
where in contrast to the $h \to b \bar b$ channel the parameter region
with $\MA \gsim 140 \gev$ is left uncovered at the $3\,\si$ level. 
This behaviour is a
consequence of the fact that the sensitivity for a SM Higgs boson 
of the same mass drops below the $3\,\si$ level when going from the 
$h \to b \bar b$ channel to the $\tau^+\tau^-$ channel.
The $5\,\si$ contours delimit the areas between the corner of 
$\MA \lsim 125 \gev$ and $\tb \gsim 35$ for the \sixoo, and $\MA \lsim 135 
\gev$ and $\tb \gsim 12$ for the \sixoooeff\ luminosity scenario. 
Correspondingly for the $3\,\si$ contours, the corner for the \sixoo\
luminosity scenario
extends up to $\MA \lsim 130 \gev$ and $\tb \gsim 25$, while the area
for the \sixoooeff\ luminosity scenario extends down to $\tb \gsim 9$. 
When compared to the results in \citere{diffH} one can see that the
regions covered by the new $5\,\si$ and 
$3\,\si$ contours are larger, but some parts are meanwhile excluded by the
Tevatron Higgs boson searches
(with the details given in the discussion of the 
$h \to b \bar b$ channel).

%%%%%%%%%%%%%%%%%%%%%%%%%%%%%%%% Begin FIGURE %%%%%%%%%%%%%%%%%%%%%%%%%%%%%%%%%
%%%%%%%%%%%%%%%%%%%%%%%%%%%%%%%% End FIGURE %%%%%%%%%%%%%%%%%%%%%%%%%%%%%%%%%%%

\smallskip
We now turn to the prospects for producing the heavier $\cp$-even Higgs 
boson of the MSSM in CED channels. The discovery reach in the 
``conventional'' search channels at the LHC, in particular 
$b\bar b \to H/A \to \tau^+\tau^-$, covers the parameter region of high
$\tb$ and not too large $\MA$~\cite{atlastdr,atlasrev,cms,CMS-TDR}, 
while a ``wedge region''~\cite{atlastdr,CMS-TDR,higgscms} remains
where the heavy MSSM Higgs bosons escape detection at the LHC
(the discovery reach is somewhat extended if decays of the heavy MSSM
Higgs bosons into supersymmetric particles can be
utilized~\cite{atlastdr,CMS-TDR}). CED production of the heavier
$\cp$-even Higgs boson of the MSSM with subsequent decay into bottom
quarks provides a unique opportunity for accessing its bottom Yukawa
coupling in a mass range where for a SM Higgs boson the decay rate into
bottom quarks would be negligibly small. 

It is well known that the properties of the heavier $\cp$-even Higgs boson
of the MSSM differ very significantly from the ones of a SM Higgs in 
the region where $\MH, \MA \gsim 150 \gev$. While for a SM Higgs the 
$\br(H \to b \bar b)$ is strongly suppressed in this mass region, the 
decay into bottom quarks is the dominant decay mode for the heavier
$\cp$-even MSSM Higgs boson (as long as no decays into supersymmetric
particles or lighter Higgs bosons are open). 
In \citere{diffH} it was shown that the MSSM process in the 
$H \to b \bar b$ channel is significantly enhanced compared to the 
SM case essentially everywhere in the unexcluded part of the 
$\MA$--$\tb$ plane. Here also the $\db$ corrections play a significant
role~\cite{benchmark3}. 
While in the case of positive $\mu$ the higher-order contribution $\db$
yields a suppression of the bottom Yukawa coupling, the opposite effect
occurs if the parameter $\mu$ is negative. 

\begin{figure}[htb!]
%\vspace{0.5em}
\begin{center}
\includegraphics[width=14cm,height=8.8cm]
                {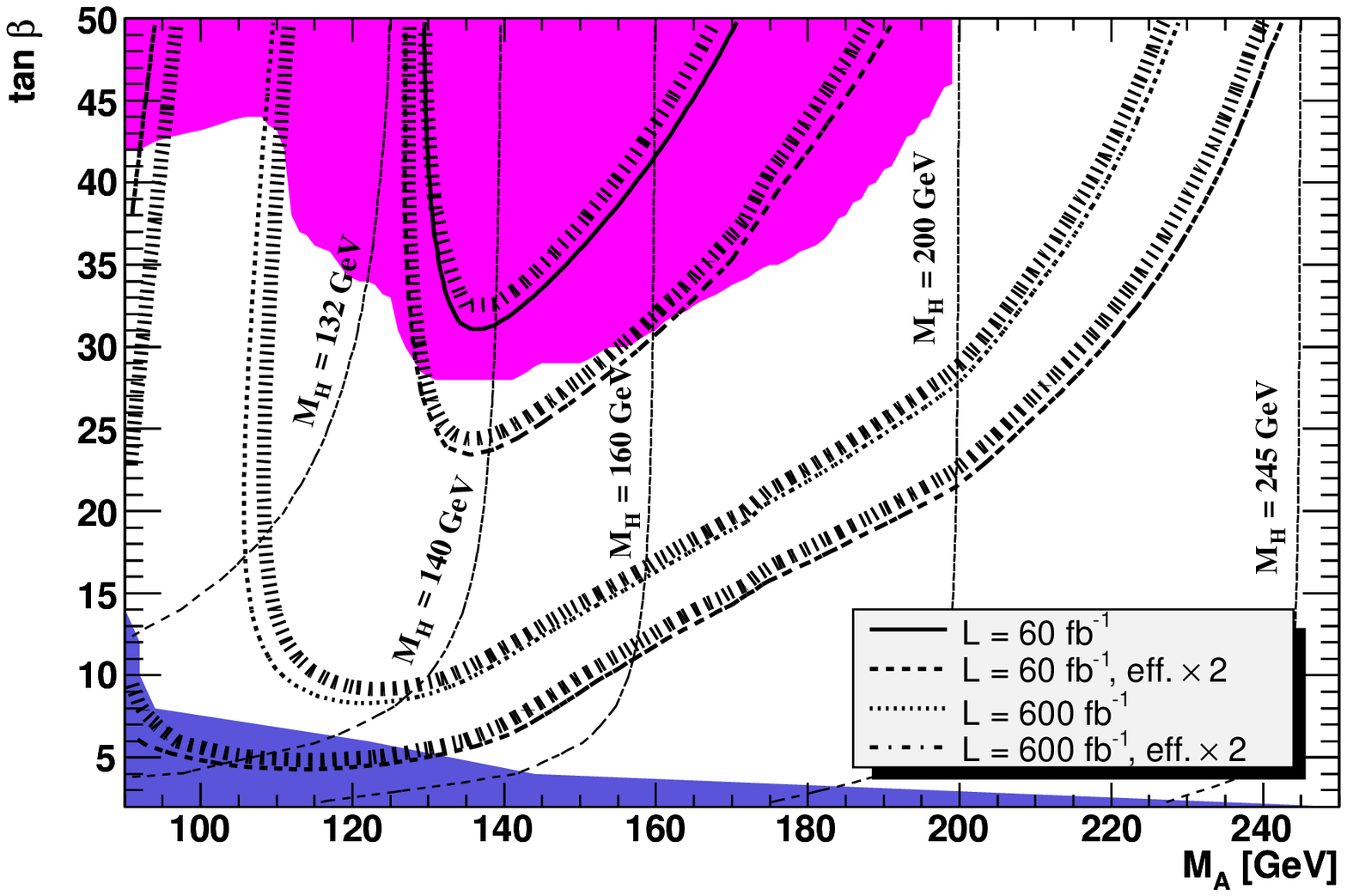}
\includegraphics[width=14cm,height=8.8cm]
                {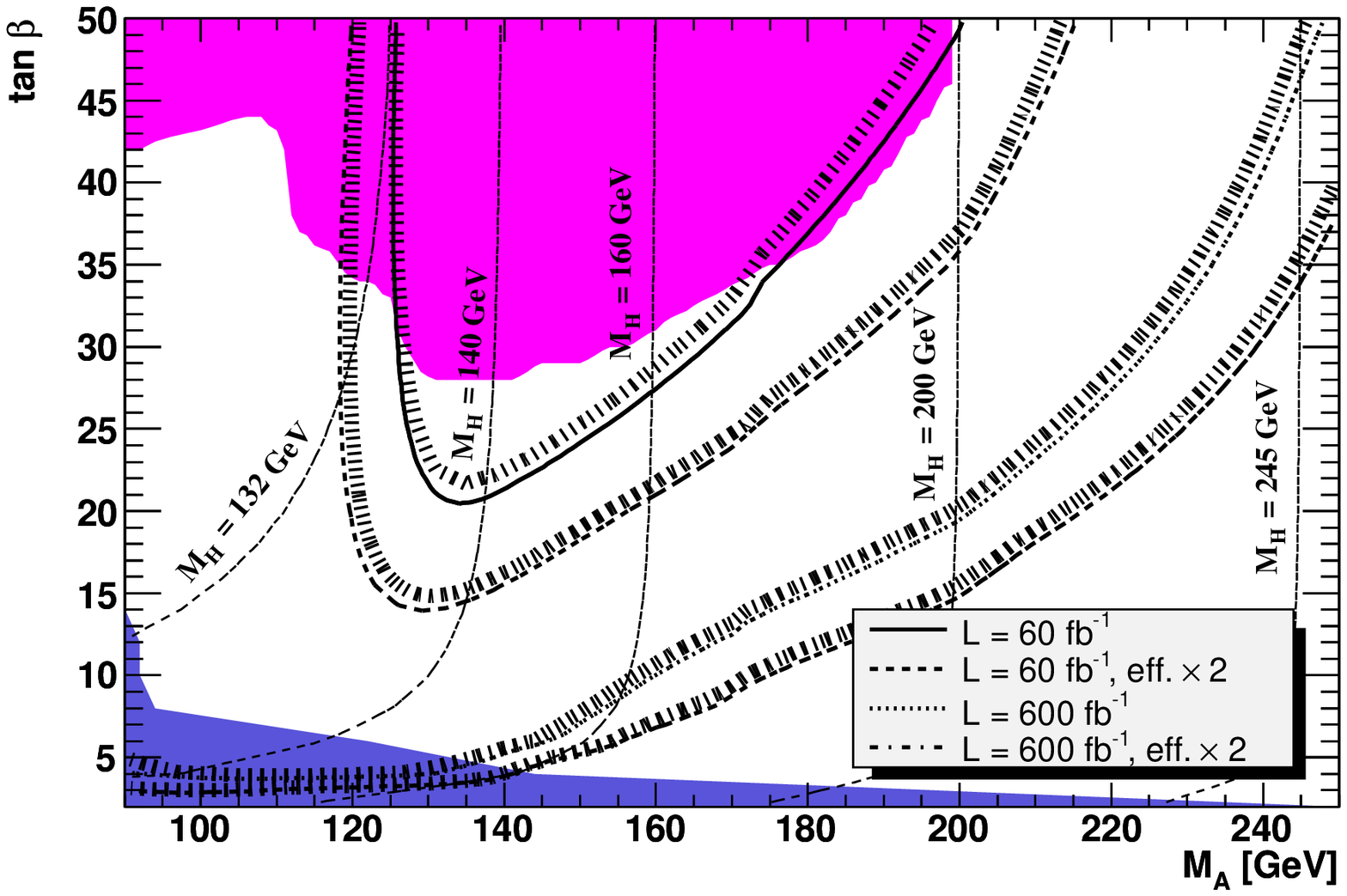}
%\vspace{1em}
\caption{
$5\,\si$ discovery contours (upper plot) and contours of $3\,\si$
statistical significance (lower plot) for the $H \to b \bar b$ channel in
CED production in the $\MA$--$\tb$ plane of the MSSM within the $\Mhmax$
benchmark scenario. The results are shown for assumed effective
luminosities (see text, combining ATLAS and CMS) of \sixoo,
\sixooeff, \sixooo\ and \sixoooeff. 
The values of $\MH$ are
indicated by contour lines. The dark shaded (blue)
region corresponds to the parameter region that
is excluded by the LEP Higgs 
searches, the lighter shaded (pink) areas are excluded by Tevatron Higgs
searches. 
}
\label{fig:Hbb1}
\end{center}
\end{figure}

\begin{figure}[htb!]
%\vspace{0.5em}
\begin{center}
\includegraphics[width=14cm,height=8.8cm]
                {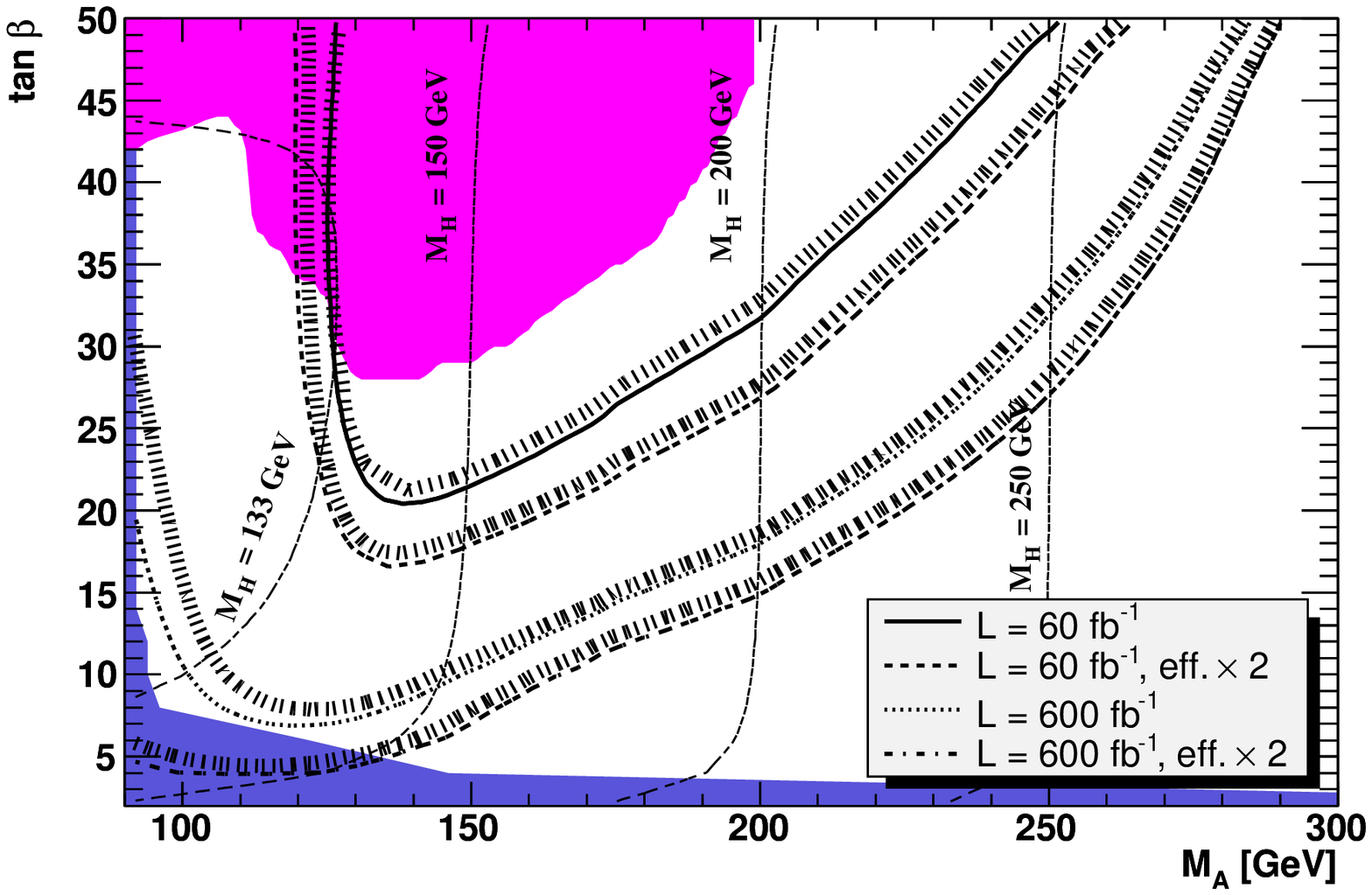}
\includegraphics[width=14cm,height=8.8cm]
                {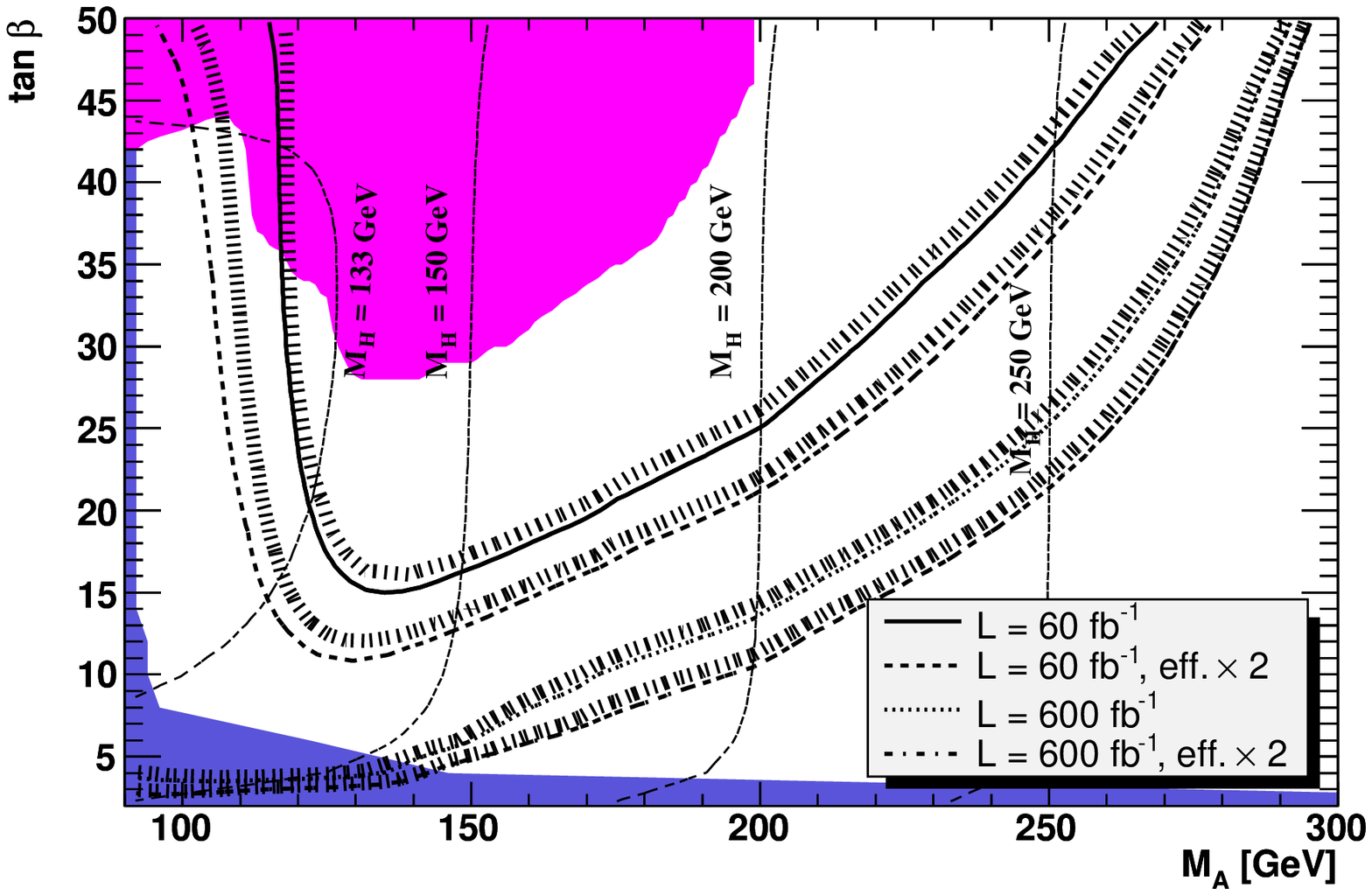}
%\vspace{1em}
\caption{
$5\,\si$ discovery contours (upper plot) and contours of $3\,\si$
statistical significance (lower plot) for the $H \to b \bar b$ channel in
CED production in the $\MA$--$\tb$ plane of the MSSM within the $\Mhmax$
benchmark scenario, but with $\mu = -500 \gev$. 
The results are shown for assumed effective
luminosities (see text, combining ATLAS and CMS) of \sixoo,
\sixooeff, \sixooo\ and \sixoooeff. 
The values of $\MH$ are
indicated by contour lines. The dark shaded (blue)
region corresponds to the parameter region that
is excluded by the LEP Higgs 
searches, the lighter shaded (pink) areas are excluded by Tevatron Higgs
searches. 
}
\label{fig:Hbb1m}
\end{center}
\end{figure}

\begin{figure}[htb!]
%\vspace{0.5em}
\begin{center}
\includegraphics[width=14cm,height=8.8cm]
                {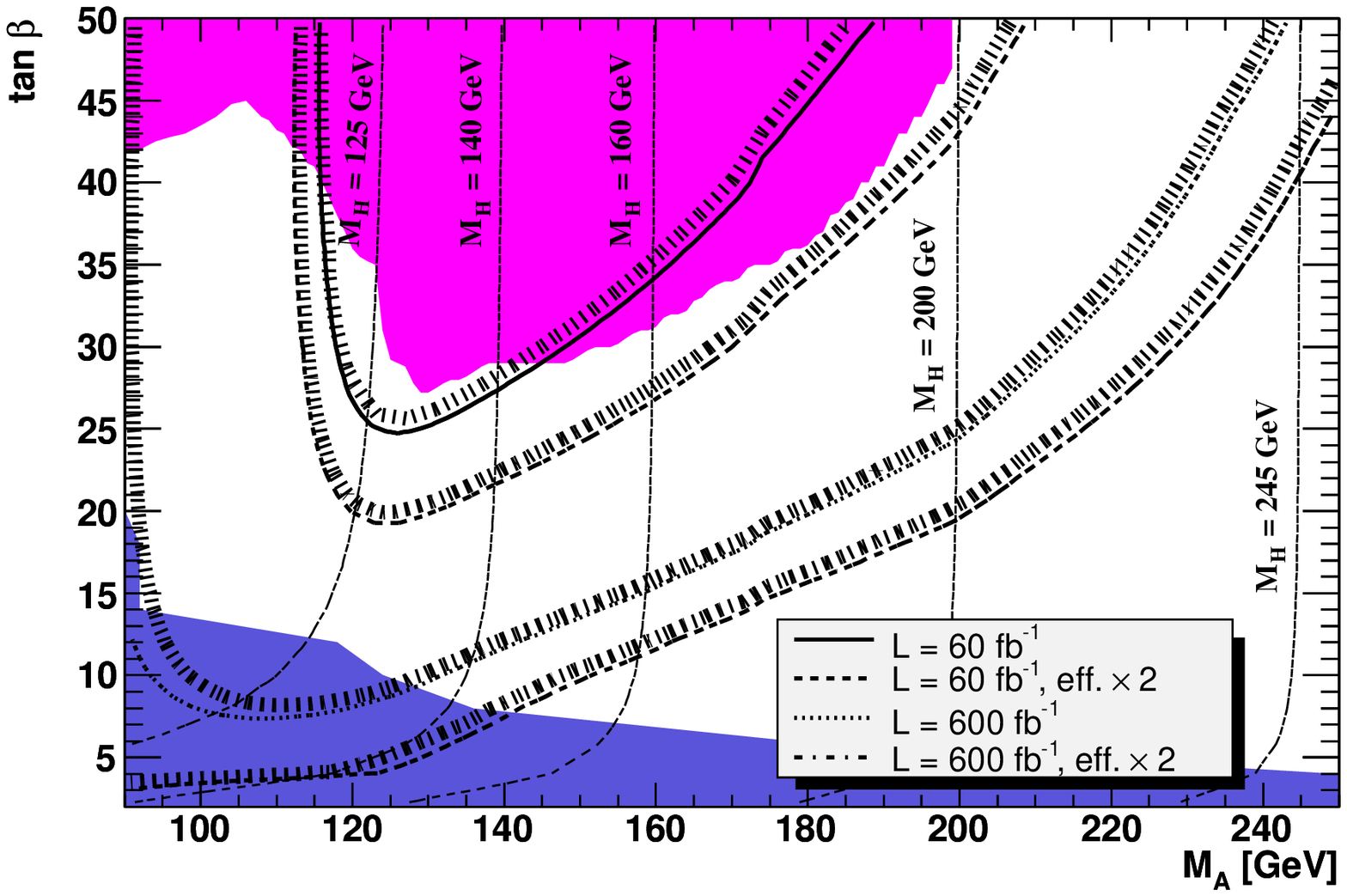}
\includegraphics[width=14cm,height=8.8cm]
                {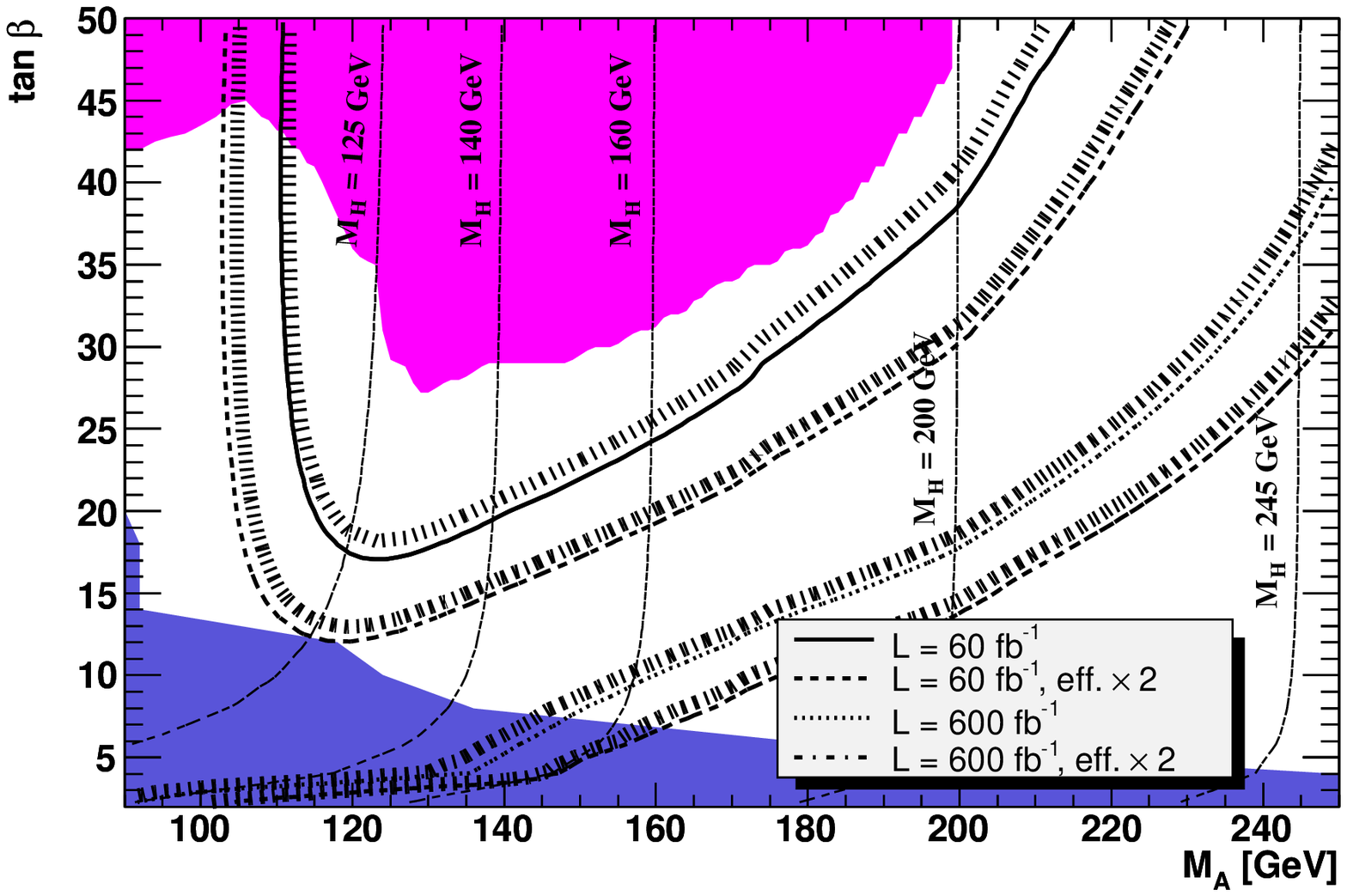}
%\vspace{1em}
\caption{
$5\,\si$ discovery contours (upper plot) and contours of $3\,\si$
statistical significance (lower plot) for the $H \to b \bar b$ channel in
CED production in the $\MA$--$\tb$ plane of the MSSM within the no-mixing
benchmark scenario. The results are shown for assumed effective
luminosities (see text, combining ATLAS and CMS) of \sixoo,
\sixooeff, \sixooo\ and \sixoooeff. 
The values of $\MH$ are
indicated by contour lines. The dark shaded (blue)
region corresponds to the parameter region that
is excluded by the LEP Higgs 
searches, the lighter shaded (pink) areas are excluded by Tevatron Higgs
searches. 
}
\label{fig:Hbb2}
\end{center}
\end{figure}

In \reffis{fig:Hbb1}--\ref{fig:Hbb2} we show the contours of $5\,\si$ 
(upper plots) and $3\,\si$ (lower plots) statistical significances
obtained in the four luminosity scenarios specified above for
the $H \to b \bar b$ channel in CED production within the $\Mhmax$
scenario for $\mu = +200\gev$, the $\Mhmax$ scenario for 
$\mu = -500\gev$ and the no-mixing scenario for $\mu = +200\gev$,
respectively. 
According to our brief discussion above (see \citeres{diffH,benchmark3}
for more details)
the following 
general pattern of the $5\,\si$ and $3\,\si$ contours is found:
the coverage in the $\MA$--$\tb$ plane is largest in
the $\Mhmax$ scenario for $\mu = -500\gev$ (\reffi{fig:Hbb1m}),
followed by the no-mixing scenario for $\mu = +200\gev$
(\reffi{fig:Hbb2}) and the $\Mhmax$ scenario for $\mu = +200\gev$
(\reffi{fig:Hbb1}). 
In the case of the lowest luminosity scenario, \sixoo, the areas given by 
the $5\,\si$ contours are largely ruled out by Tevatron Higgs-boson
searches, except for the case of $\mu = -500\gev$.
However, at the $3\,\si$ level for $\mu = -500 \gev$
(\reffi{fig:Hbb1m}) and in the no-mixing scenario (\reffi{fig:Hbb2}) 
CED production of the heavy $\cp$-even Higgs with \sixoo\ would cover 
substantial parts of the MSSM parameter space that are unexcluded by the
Higgs searches carried out so far at the Tevatron (and elsewhere).

The high-luminosity scenarios could allow a $5\,\si$
significance for $\MH \lsim 240 \gev$ for large $\tb$ in the $\Mhmax$
scenario with $\mu = +200 \gev$. This is extended to $\MH \lsim 290 \gev$
for $\mu = -500 \gev$, and reaches $\MH \lsim 255 \gev$ in the no-mixing
scenario, again with $\mu = +200 \gev$. The $3\,\si$ contours extend the
reach by up to $\sim 20 \gev$ in $\MH$. At low $\tb$ the
$3\,\si$ reach goes down to (i.e.\ crosses the LEP exclusion
bound) $\MH \sim 140-160 \gev$, depending on the scenario. 

We also note that the region left uncovered at the $3\,\si$ level
in the light $\cp$-even Higgs-boson 
analyses can to a large extent be 
covered by searches for the heavy $\cp$-even 
Higgs boson if the CED channel can be utilized 
at high instantaneous luminosity. 
Thus, a combined $h$- and $H$-analysis in this case would 
yield a reasonable signal of CED Higgs-boson production
at the $3\,\si$ level in 
nearly the whole parameter space that is unexcluded
by the LEP and Tevatron searches.

In comparison with \citere{diffH} one can see that the various updates 
(see \refse{sec:update}), as in the case of the light $\cp$-even Higgs boson, 
lead to a somewhat larger coverage in the $\MA$--$\tb$ plane.
More specifically, on average, the results are enhanced by a factor of
$\sim 1.6$ in the regions where old significances are larger than~1. 
In the rest of the parameter space they do not exceed~2.1.

To summarize, the CED production at the LHC may be
a unique way to access the bottom Yukawa coupling of a Higgs boson as
heavy as $290 \gev$ (which would obviously be a clear sign of physics
beyond the Standard Model).
Within the MSSM, however, this channel does not cover additional
parameter space of the so-called ``LHC wedge'' (see the discussion in
\citere{diffH}).

%%%%%%%%%%%%%%%%%%%%%%%%%%%%%%%% Begin FIGURE %%%%%%%%%%%%%%%%%%%%%%%%%%%%%%%%%
%%%%%%%%%%%%%%%%%%%%%%%%%%%%%%%% End FIGURE %%%%%%%%%%%%%%%%%%%%%%%%%%%%%%%%%%%

%%%%%%%%%%%%%%%%%%%%%%%%%%%%%%%% Begin FIGURE %%%%%%%%%%%%%%%%%%%%%%%%%%%%%%%%%
%%%%%%%%%%%%%%%%%%%%%%%%%%%%%%%% End FIGURE %%%%%%%%%%%%%%%%%%%%%%%%%%%%%%%%%%%

%%%%%%%%%%%%%%%%%%%%%%%%%%%%%%%% Begin FIGURE %%%%%%%%%%%%%%%%%%%%%%%%%%%%%%%%%
%%%%%%%%%%%%%%%%%%%%%%%%%%%%%%%% End FIGURE %%%%%%%%%%%%%%%%%%%%%%%%%%%%%%%%%%%

%%%%%%%%%%%%%%%%%%%%%%%%%%%%%%%% Begin FIGURE %%%%%%%%%%%%%%%%%%%%%%%%%%%%%%%%%
\begin{figure}[htb!]
\vspace{0.5em}
\begin{center}
\includegraphics[width=14cm,height=8.5cm]
                {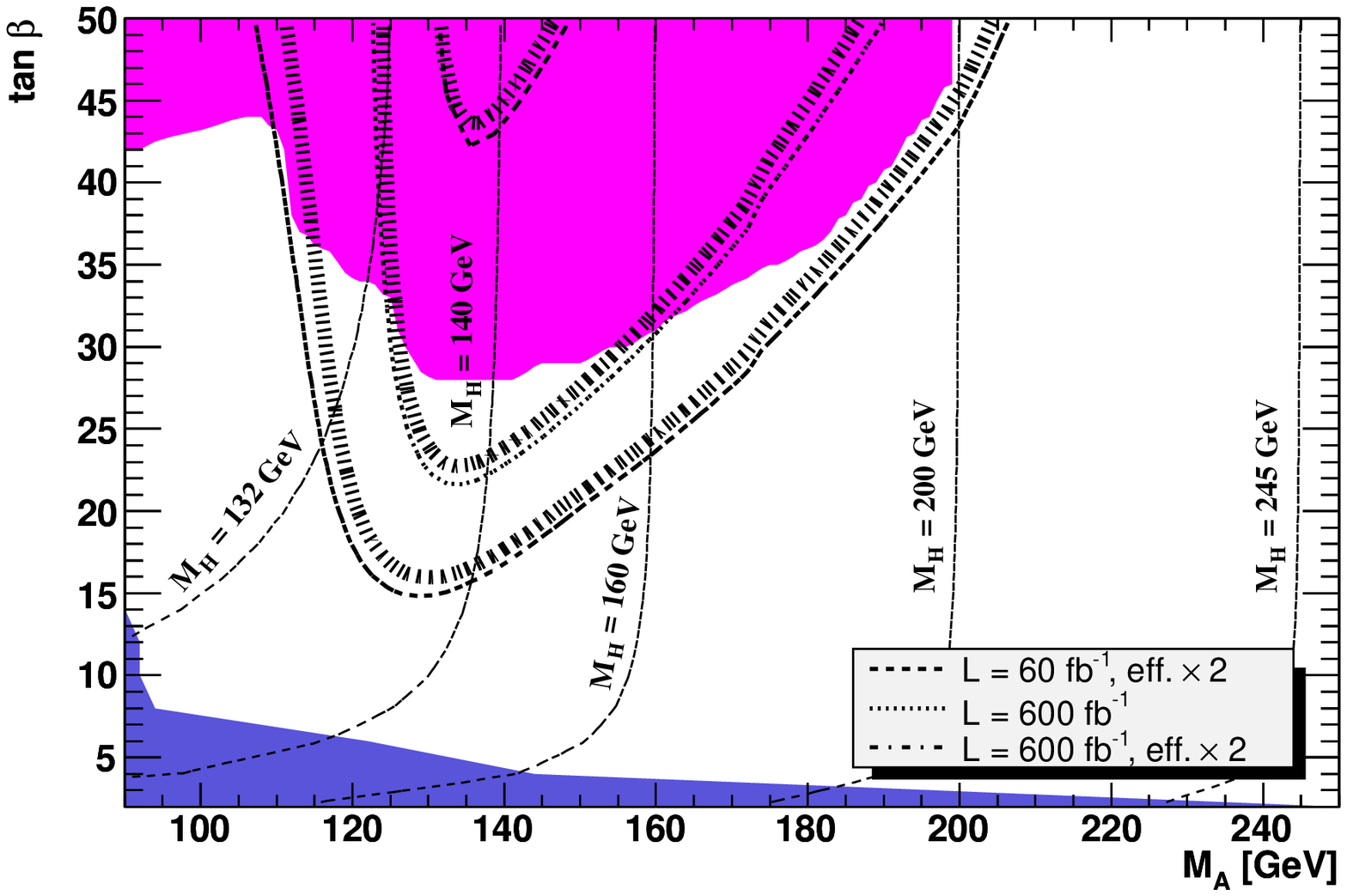}
\includegraphics[width=14cm,height=8.5cm]
                {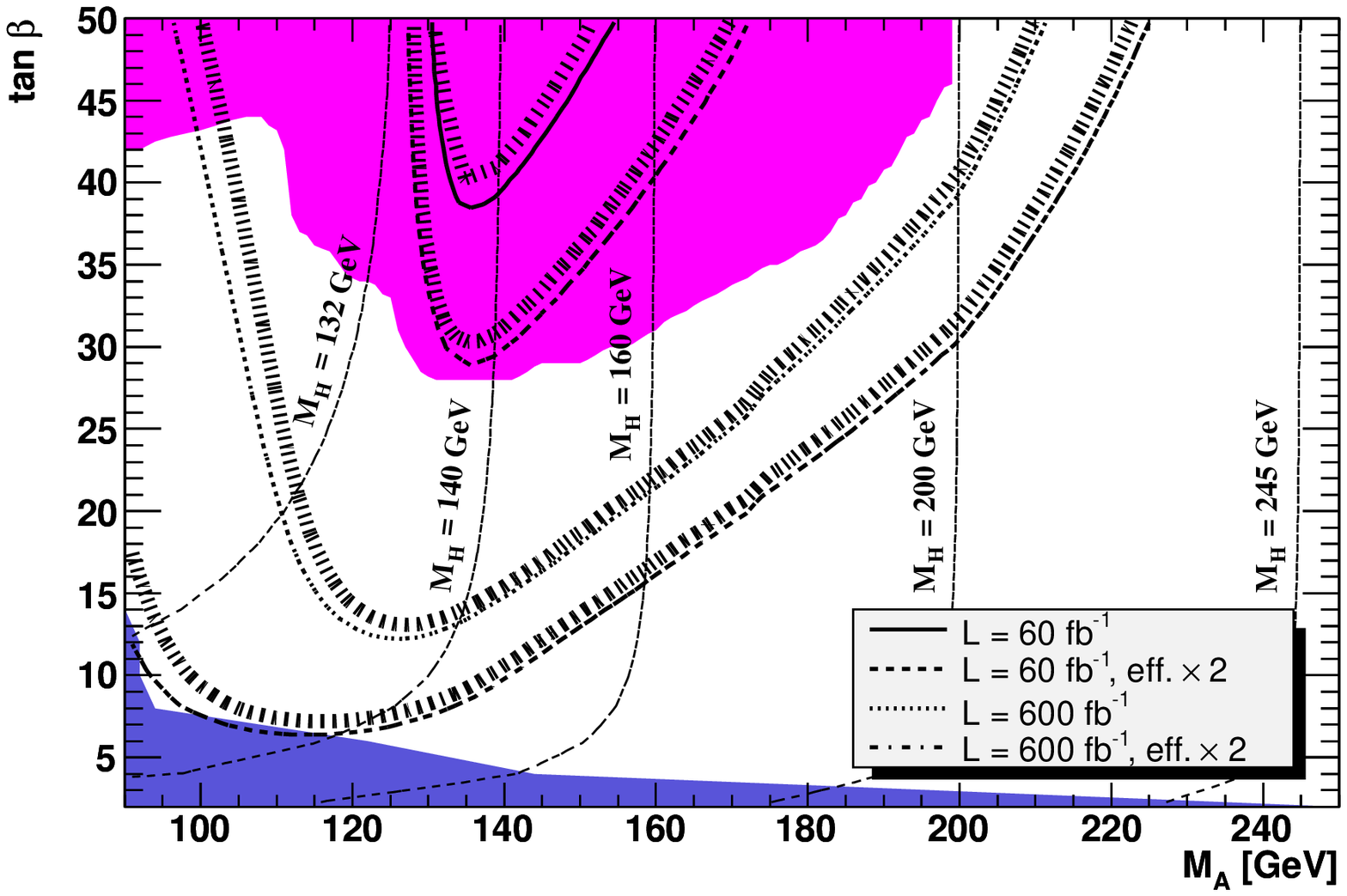}
%\vspace{1em}
\caption{
$5\,\si$ discovery contours (upper plot) and contours of $3\,\si$
statistical significance (lower plot) for the $H \to \tau^+\tau^-$ channel in
CED production in the $\MA$--$\tb$ plane of the MSSM within the $\Mhmax$
benchmark scenario. The results are shown for assumed effective
luminosities (see text, combining ATLAS and CMS) of \sixoo,
\sixooeff, \sixooo\ and \sixoooeff. 
The values of $\MH$ are
indicated by contour lines. The dark shaded (blue)
region corresponds to the parameter region that
is excluded by the LEP Higgs 
searches, the lighter shaded (pink) areas are excluded by Tevatron Higgs
searches. 
}
\label{fig:Htautau}
\end{center}
\end{figure}
%%%%%%%%%%%%%%%%%%%%%%%%%%%%%%%% End FIGURE %%%%%%%%%%%%%%%%%%%%%%%%%%%%%%%%%%%

We finish the update of the CED reach in the conventional benchmark
scenarios with the $H \to \tau^+\tau^-$ channel in the $\Mhmax$ scenario
as shown in \reffi{fig:Htautau}. The $5\,\si$ discovery regions shown
in the upper plot are only visible for the high-luminosity scenarios,
(apart from a small wedge around $\MA = 140 \gev$ and $\tb \gsim 43$,
where also lower luminosity could yield $5\,\si$).
However, they are excluded to a large extent by the Tevatron Higgs-boson
searches.  
In the \sixoooeff\ scenario, $5\,\si$ coverage is reached 
for $\tb \sim 15 -  30$ and for 
$\MA = 110 - 180 \gev$.
The $3\,\si$ areas, shown in the lower plot, extend substantially
outside the parameter space probed by the Tevatron in the high
luminosity scenarios. They cover up to 
$\MH \lsim 220 \gev$ for large $\tb$. As for the $H \to b \bar b$
channel, the reach is larger than in the
analysis of \citere{diffH}. 
More specifically, on average, the results are enhanced by a factor of
$\sim 1.4$ in the regions where old significances are larger than~1. In
the rest of the parameter space they do not exceed~1.6.

%%%%%%%%%%%%%%%%%%%%%%%%%%%%%%%%%%%%%%%%%%%%%%%%%%%%%%%%%%%%%%%%%%%%%%%%%%%%%%%
%%%%%%%%%%%%%%%%%%%%%%%%%%%%%%%%%%%%%%%%%%%%%%%%%%%%%%%%%%%%%%%%%%%%%%%%%%%%%%%

\subsection{Discovery reach for neutral $\cp$-even Higgs bosons 
in the ``CDM benchmark scenarios''}
\label{sec:cdmbench}

In this section we present for the first time a detailed study of
the prospects for 
observing the neutral $\cp$-even MSSM Higgs bosons in CED production in
the ``CDM benchmark scenarios'' \pdrei\ and \pvier, see \refse{sec:update}
(see also \citere{ismd}).
As before the results are displayed
in $\MA$--$\tb$ planes, where \pdrei\ (\pvier) is defined 
for values below  
$\tb = 40$ (50) and for $\MA \ge 100 \gev$. 
As in the previous subsection, the plots also show contour lines for
$\Mh$, $\MH$ 
as well as 
the parameter regions excluded by the LEP Higgs searches (as dark
shaded (blue) areas) and Tevatron Higgs-boson searches (as lighter
shaded (pink) areas) as obtained with {\tt HiggsBounds}~\cite{higgsbounds}.
It should be kept in mind that large parts of these planes are also in
agreement with electroweak precision data and $B$~physics data~\cite{CDM}.
In this sense these scenarios fulfill all external constraints,
contrary to the conventional $\Mhmax$ and no-mixing benchmark
scenarios (which were designed to highlight specific
characteristics of the MSSM Higgs sector).

The $5\,\si$ discovery contours as well as the contours for $3\,\si$
significances have been obtained in the same way as for the conventional
benchmark scenarios. In general they show similar qualitative
features as the results in the $\Mhmax$ and the no-mixing scenario.

\begin{figure}[htb!]
\vspace{0.5em}
\begin{center}
\includegraphics[width=14cm,height=8.5cm]
                {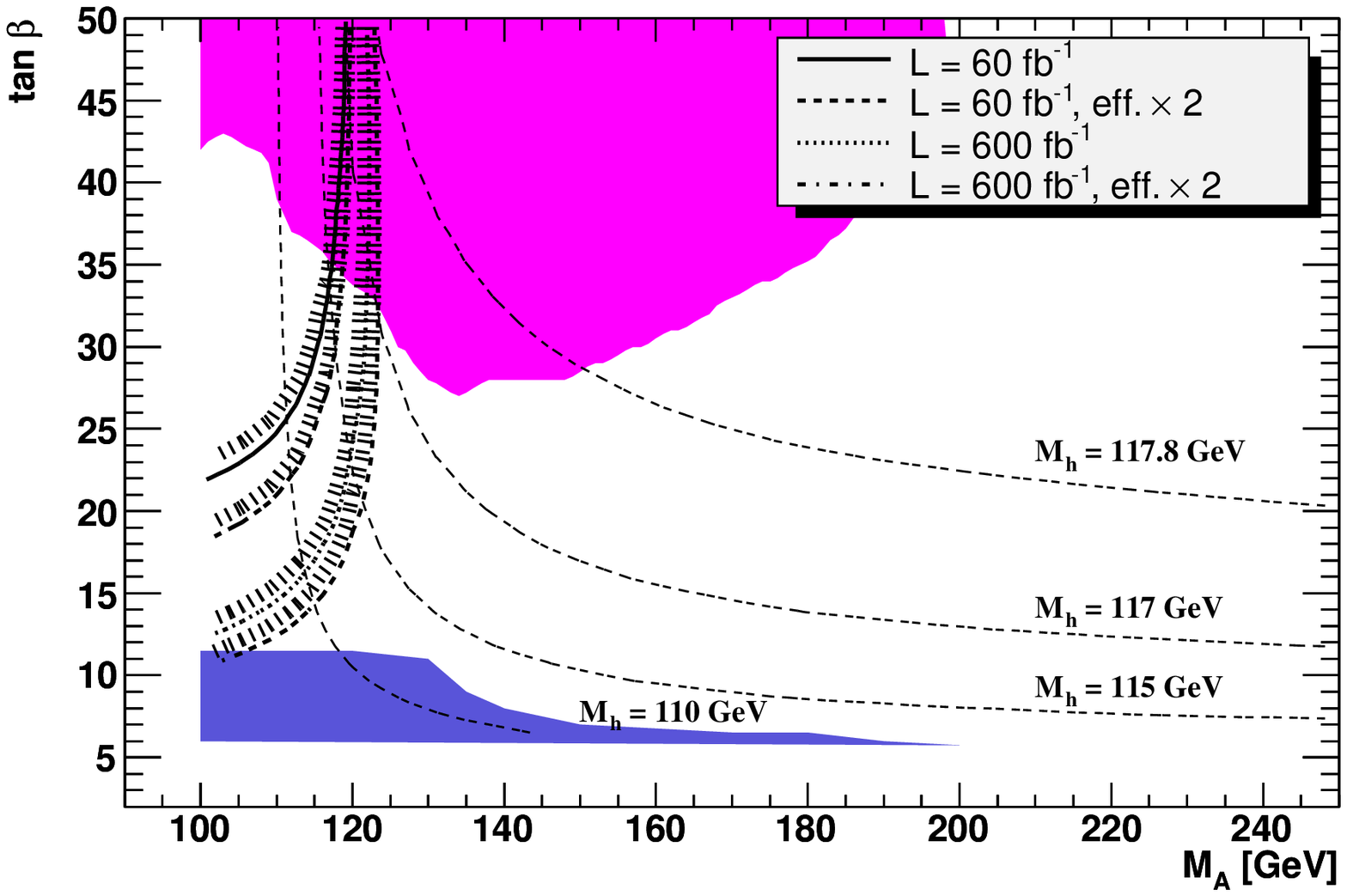}
\includegraphics[width=14cm,height=8.5cm]
                {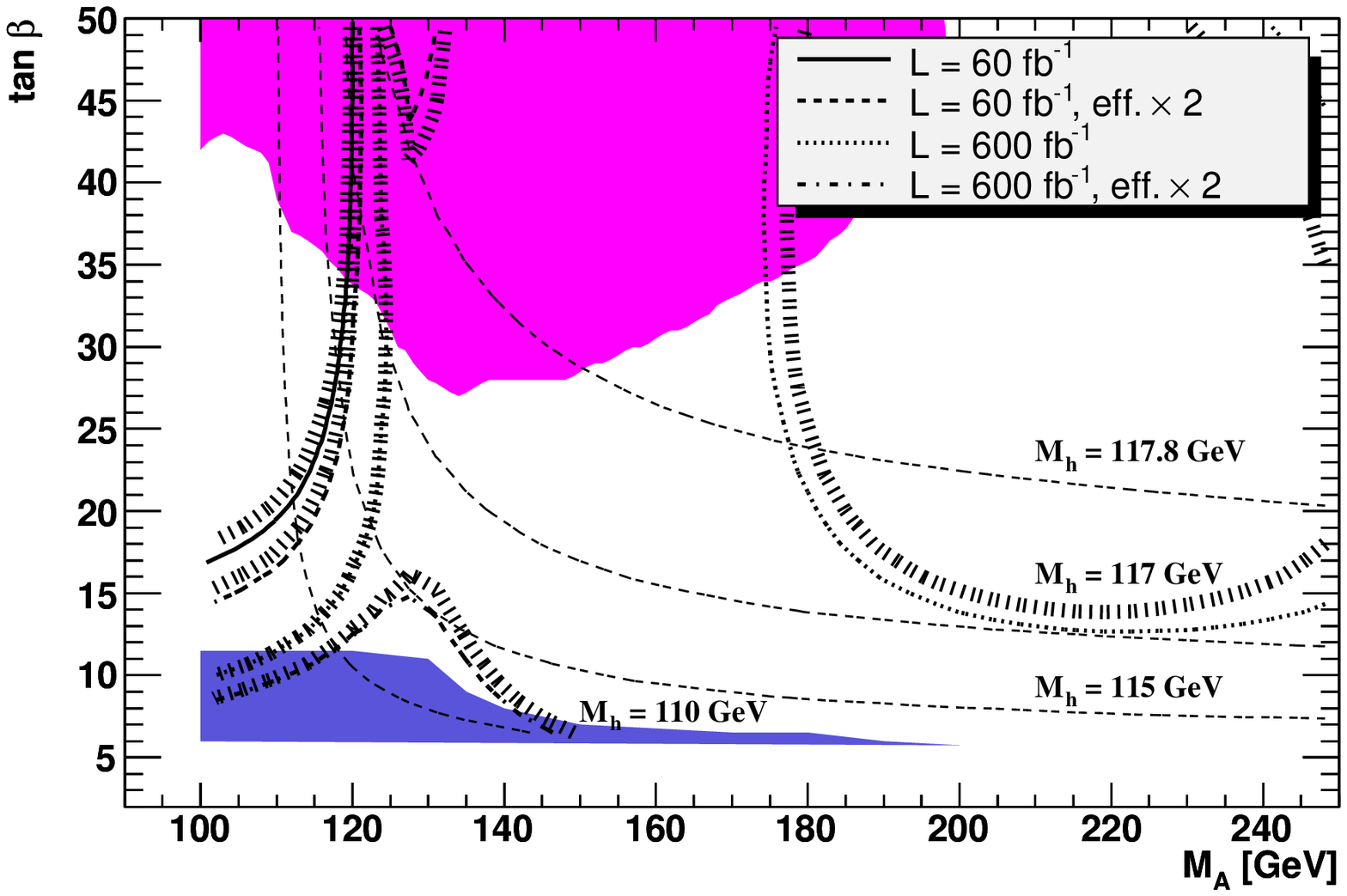}
%\vspace{1em}
\caption{
$5\,\si$ discovery contours (upper plot) and contours of $3\,\si$
statistical significance (lower plot) for the $h \to b \bar b$ channel in
CED production in the $\MA$--$\tb$ plane of the MSSM within the 
``CDM benchmark'' scenario \pdrei. The results are shown for assumed
effective luminosities (see text, combining ATLAS and CMS) of \sixoo,
\sixooeff, \sixooo\ and \sixoooeff. 
The values of $\Mh$ are
indicated by contour lines. The dark shaded (blue)
region corresponds to the parameter region that
is excluded by the LEP Higgs 
searches, the lighter shaded (pink) areas are excluded by Tevatron Higgs
searches. 
}
\label{fig:hbb-p3}
\end{center}
\end{figure}

The results for the channel $h \to b \bar b$ in \pdrei\ are presented in
\reffi{fig:hbb-p3}. The upper and lower plots show the contours for
$5\,\si$ discovery and $3\,\si$ significance, respectively. 
A $5\,\si$ discovery is possible for $\MA \lsim 125 \gev$ 
and moderate to large values of $\tb$, i.e.\ 
$\tb \gsim 10$, depending on the luminosity
scenario. At the $3\,\si$ level,
the coverage extends to the rest of the plane in the
\sixoooeff\ scenario, leaving 
only a small uncovered funnel around $\MA \approx 125 \gev$ and 
$\tb \lsim 15$ or $\tb \gsim 42$.
The LEP bounds exclude a region at $\tb \lsim 11$ and 
$\MA \lsim 140 \gev$ that is barely touched by the $5\,\si$ discovery
contours. 
However, the LEP searches exclude a large part of the funnel at 
small $\tb$ that is
left uncovered at the $3\,\si$ level by CED Higgs searches in the 
\sixoooeff\ scenario. Thus, the LEP
exclusion regions are complementary to the parameter space covered 
by CED Higgs-boson production in
\pdrei.
(As mentioned above, the fact that the LEP bounds do not reach
$\Mh = 114 \gev$~\cite{LEPHiggsSM,LEPHiggsMSSM} reflects the
theory uncertainty of $\sim 3 \gev$ in $\Mh$~\cite{mhiggsAEC}
taken into account for the LEP bounds.)
 
The Tevatron searches exclude a parameter space with $\tb \gsim 28$
around $\MA = 140 \gev$.
Consequently, the uncovered funnel at $\tb \gsim 42$ for CED Higgs-boson
production in the \sixoooeff\ scenario lies within the parameter region
that is excluded by the Tevatron Higgs searches.

\begin{figure}[htb!]
\vspace{0.5em}
\begin{center}
\includegraphics[width=14cm,height=8.5cm]
                {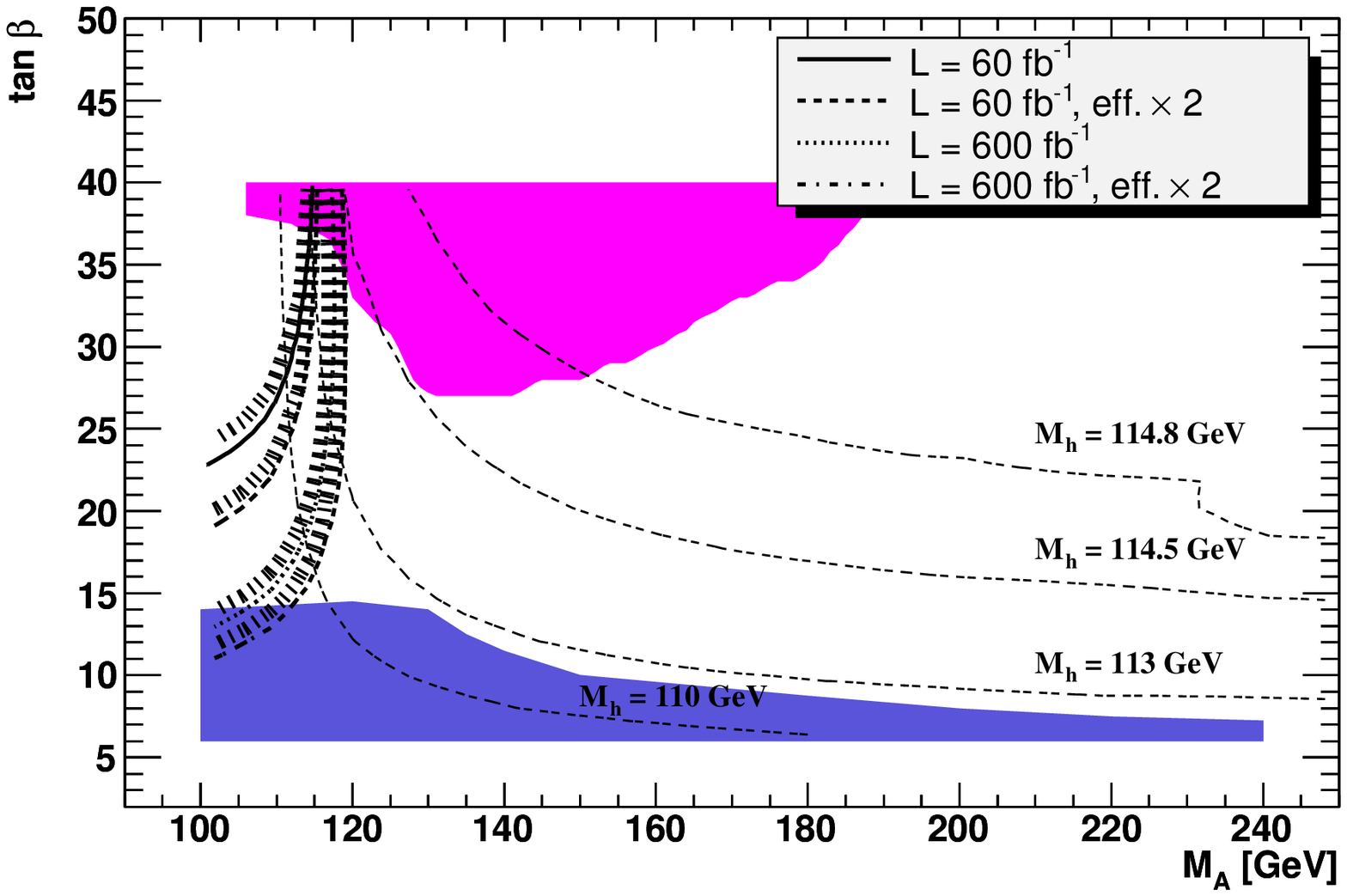}
\includegraphics[width=14cm,height=8.5cm]
                {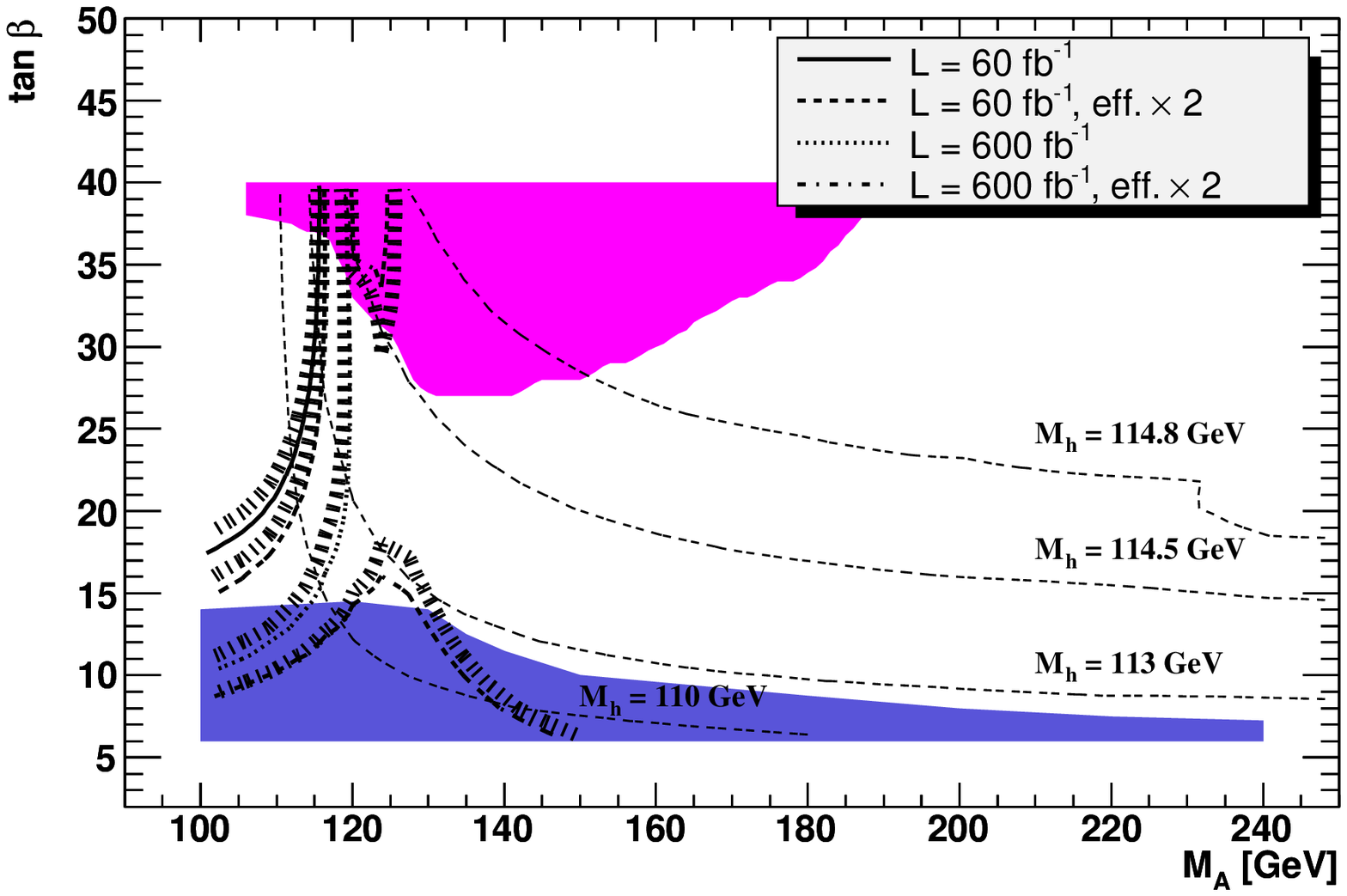}
%\vspace{1em}
\caption{
$5\,\si$ discovery contours (upper plot) and contours of $3\,\si$
statistical significance (lower plot) for the $h \to b \bar b$ channel in
CED production in the $\MA$--$\tb$ plane of the MSSM within the 
``CDM benchmark'' scenario \pvier. The results are shown for assumed
effective luminosities (see text, combining ATLAS and CMS) of \sixoo,
\sixooeff, \sixooo\ and \sixoooeff. 
The values of $\Mh$ are
indicated by contour lines. The dark shaded (blue)
region corresponds to the parameter region that
is excluded by the LEP Higgs 
searches, the lighter shaded (pink) areas are excluded by Tevatron Higgs
searches. 
}
\label{fig:hbb-p4}
\end{center}
\end{figure}

In \reffi{fig:hbb-p4} we present the corresponding results in the 
$h \to b \bar b$ channel for \pvier, which, as explained above, extends
only up to $\tb = 40$. The $5\,\si$ and $3\,\si$ reach is similar to that in
\pdrei. However, due to the lower $\Mh$ values realized in this scenario
the LEP searches exclude an even larger part of the funnel at 
small $\tb$ that is
left uncovered at the $3\,\si$ level by CED Higgs searches in the 
\sixoooeff\ scenario.

%%%%%%%%%%%%%%%%%%%%%%%%%%%%%%%% Begin FIGURE %%%%%%%%%%%%%%%%%%%%%%%%%%%%%%%%%
%%%%%%%%%%%%%%%%%%%%%%%%%%%%%%%% End FIGURE %%%%%%%%%%%%%%%%%%%%%%%%%%%%%%%%%%%

%%%%%%%%%%%%%%%%%%%%%%%%%%%%%%%% Begin FIGURE %%%%%%%%%%%%%%%%%%%%%%%%%%%%%%%%%
%%%%%%%%%%%%%%%%%%%%%%%%%%%%%%%% End FIGURE %%%%%%%%%%%%%%%%%%%%%%%%%%%%%%%%%%%

%%%%%%%%%%%%%%%%%%%%%%%%%%%%%%%% Begin FIGURE %%%%%%%%%%%%%%%%%%%%%%%%%%%%%%%%%
\begin{figure}[htb!]
\vspace{0.5em}
\begin{center}
\includegraphics[width=14cm,height=8.5cm]
                {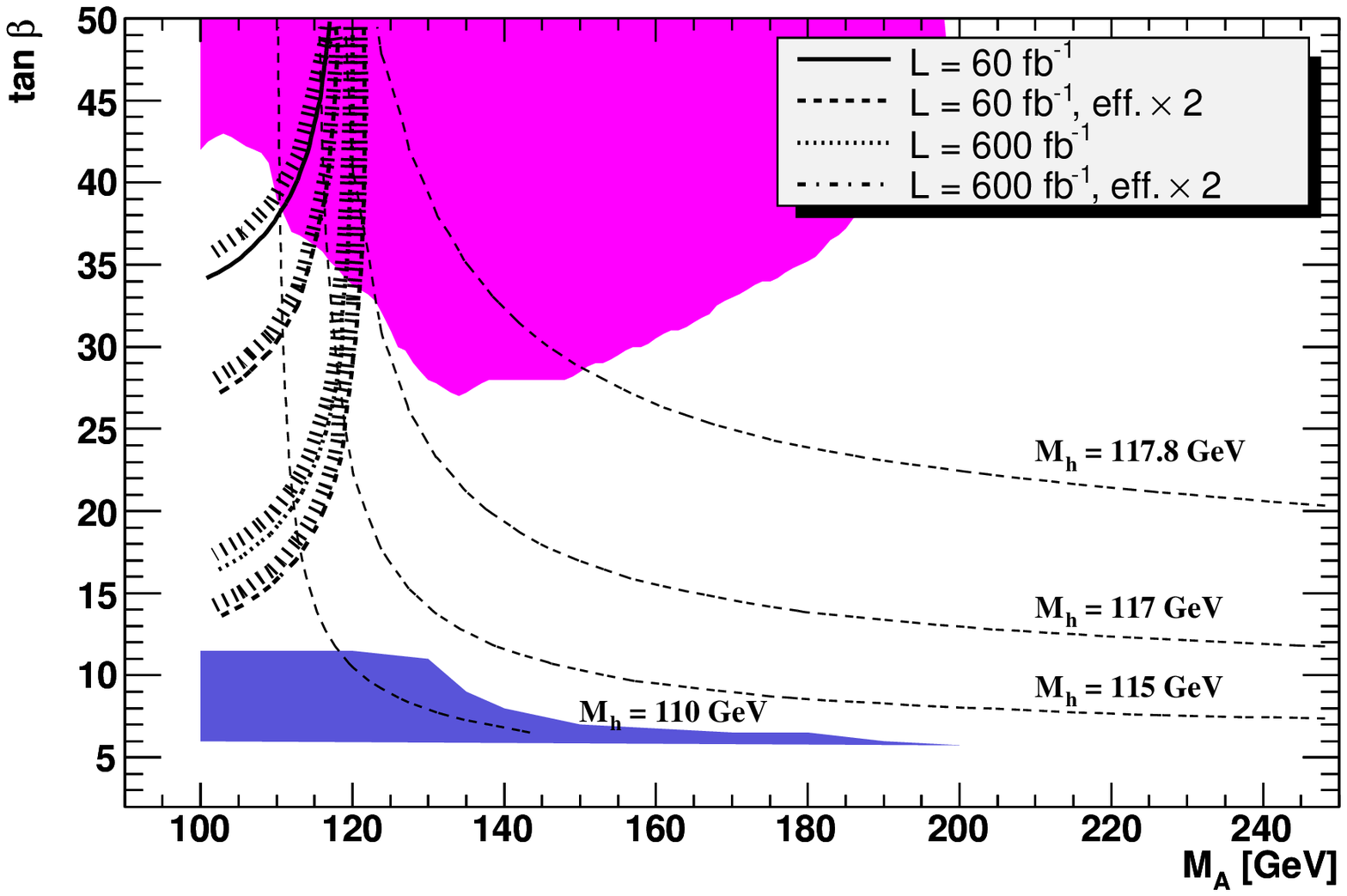}
\includegraphics[width=14cm,height=8.5cm]
                {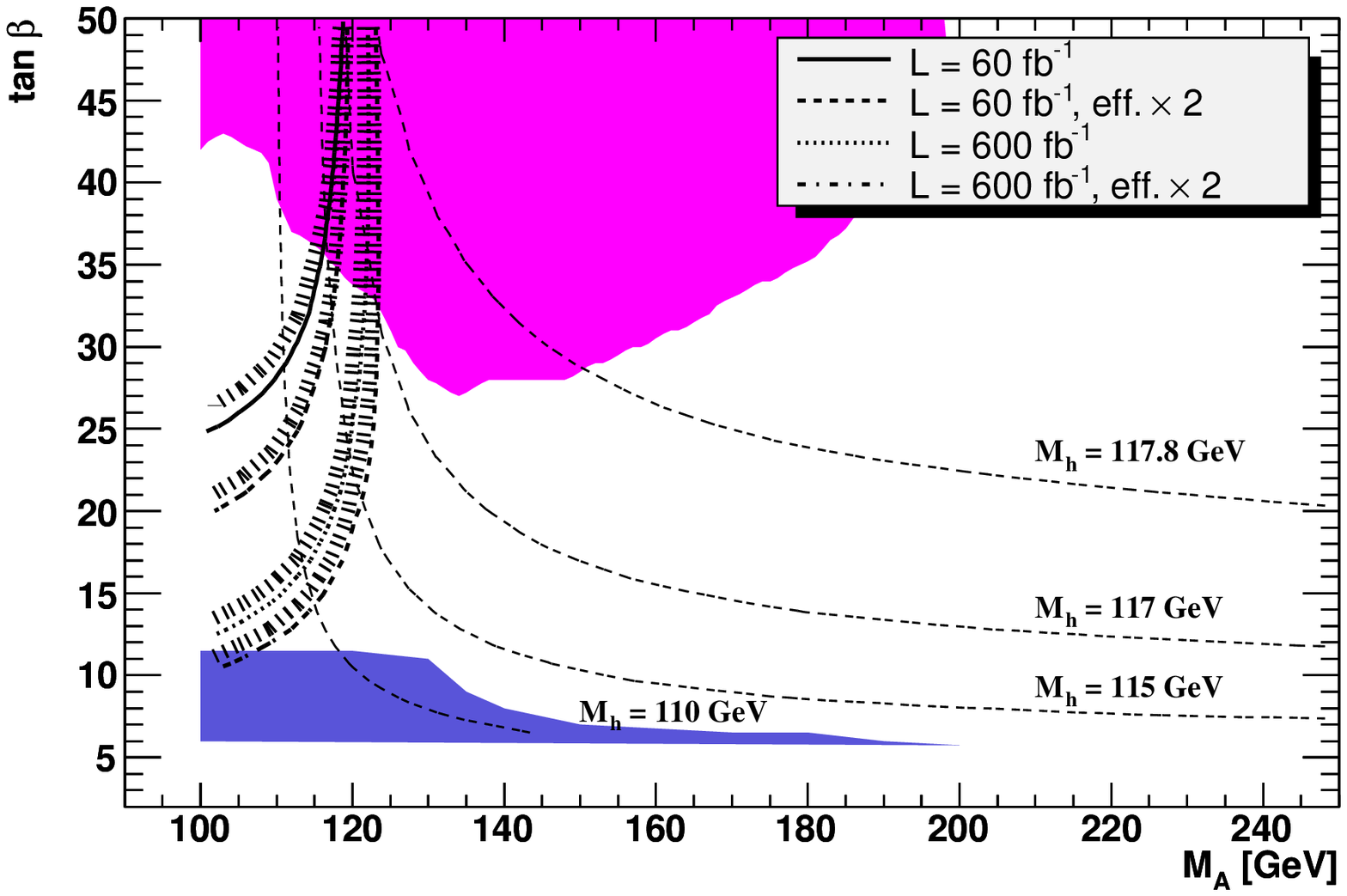}
%\vspace{1em}
\caption{
$5\,\si$ discovery contours (upper plot) and contours of $3\,\si$
statistical significance (lower plot) for the $h \to \tau^+\tau^-$ channel in
CED production in the $\MA$--$\tb$ plane of the MSSM within the 
``CDM benchmark'' scenario \pdrei. The results are shown for assumed
effective luminosities (see text, combining ATLAS and CMS) of \sixoo,
\sixooeff, \sixooo\ and \sixoooeff. 
The values of $\Mh$ are
indicated by contour lines. The dark shaded (blue)
region corresponds to the parameter region that
is excluded by the LEP Higgs 
searches, the lighter shaded (pink) areas are excluded by Tevatron Higgs
searches. 
}
\label{fig:htautau-p3}
\end{center}
\end{figure}
%%%%%%%%%%%%%%%%%%%%%%%%%%%%%%%% End FIGURE %%%%%%%%%%%%%%%%%%%%%%%%%%%%%%%%%%%

In \reffi{fig:htautau-p3} we show the results for the
$h \to \tau^+\tau^-$ channel in the scenario \pdrei. For the high luminosity
cases a $5\,\si$ coverage is obtained for
$\tb \gsim 15$ and $\MA \lsim 120 \gev$. A slightly larger region is
covered at the $3\,\si$ level, extending 
down to $\tb \gsim 10$ in the high luminosity
scenarios. Compared with the $\Mhmax$ scenario, the coverage in $\MA$ is
smaller by $\sim 10 \gev$.

\begin{figure}[htb!]
\vspace{0.5em}
\begin{center}
\includegraphics[width=14cm,height=8.5cm]
                {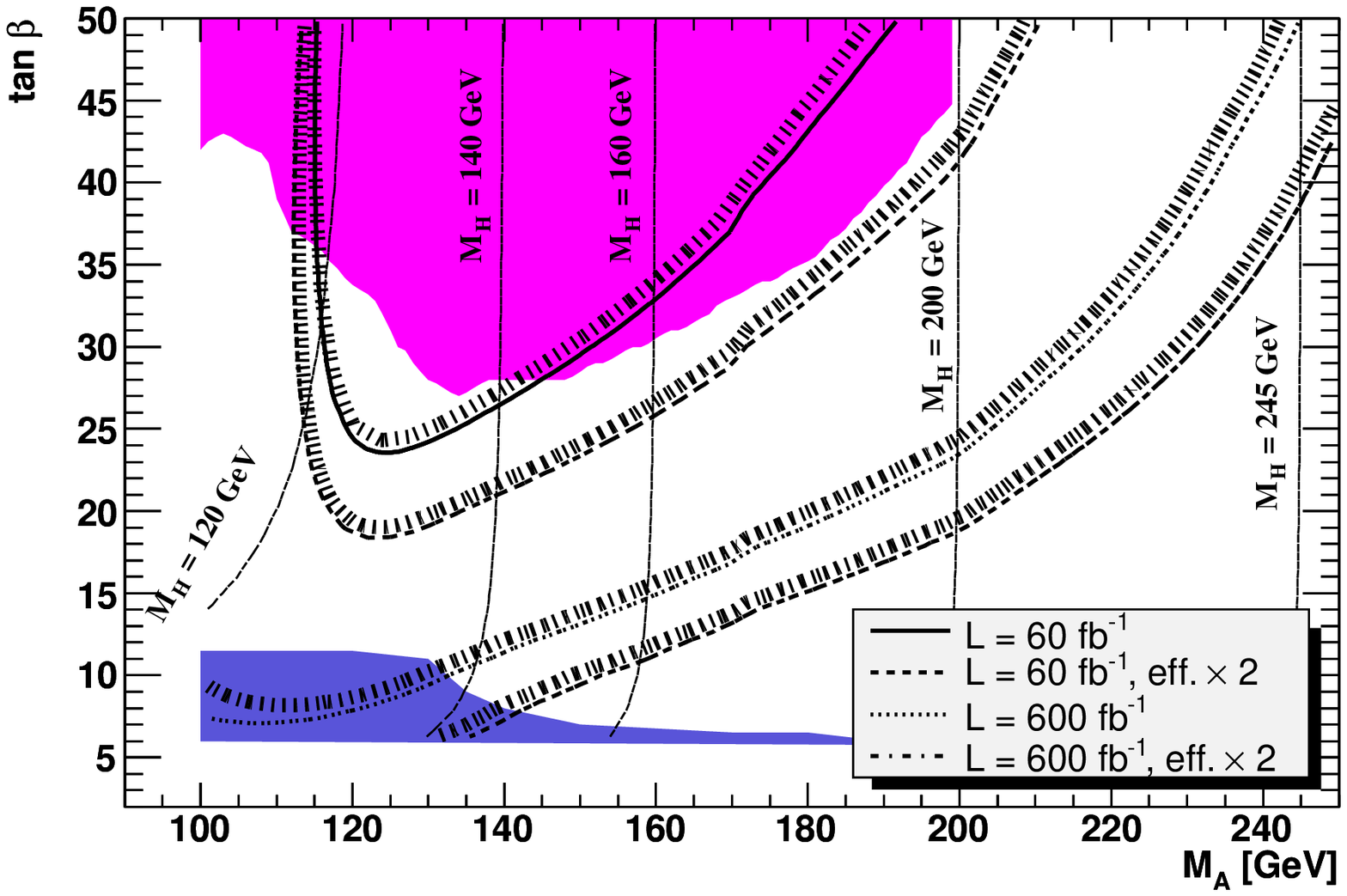}
\includegraphics[width=14cm,height=8.5cm]
                {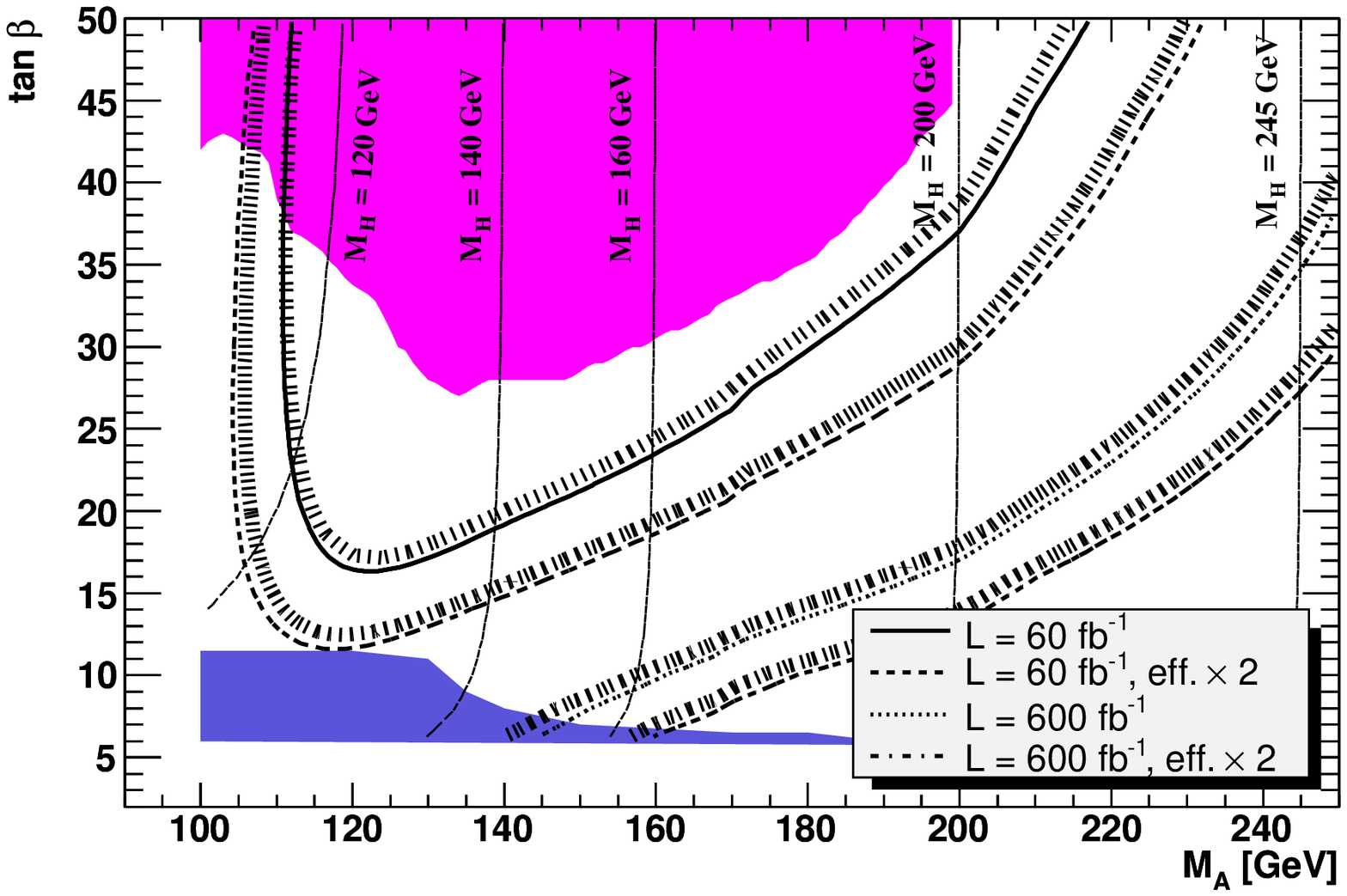}
%\vspace{1em}
\caption{
$5\,\si$ discovery contours (upper plot) and contours of $3\,\si$
statistical significance (lower plot) for the $H \to b \bar b$ channel in
CED production in the $\MA$--$\tb$ plane of the MSSM within the 
``CDM benchmark'' scenario \pdrei. The results are shown for assumed
effective luminosities (see text, combining ATLAS and CMS) of \sixoo,
\sixooeff, \sixooo\ and \sixoooeff. 
The values of $\MH$ are
indicated by contour lines. The dark shaded (blue)
region corresponds to the parameter region that
is excluded by the LEP Higgs 
searches, the lighter shaded (pink) areas are excluded by Tevatron Higgs
searches. 
}
\label{fig:Hbb-p3}
\end{center}
\end{figure}

\smallskip
Now we turn to the analysis for the heavy $\cp$-even Higgs boson. 
In \reffi{fig:Hbb-p3} we present the results for the channel 
$H \to b \bar b$ in the scenario \pdrei. The reach is somewhat larger
than in the $\Mhmax$ scenario and similar to the no-mixing scenario. 
$5\,\si$ can be reached up to $\MH \lsim 260 \gev$ at large $\tb$ and
high luminosity. At low luminosity the region extends only up to 
$\MH \lsim 210 \gev$, and it is largely excluded by the Tevatron
searches. The $3\,\si$ significance contours move to larger $\MH$ values
by $20-30 \gev$. As above, the region
left uncovered at the $3\,\si$ level by CED light Higgs searches in the 
\sixoooeff\ scenario is accessible via CED
production of the heavy $\cp$-even Higgs boson. 

\begin{figure}[htb!]
\vspace{0.5em}
\begin{center}
\includegraphics[width=14cm,height=8.5cm]
                {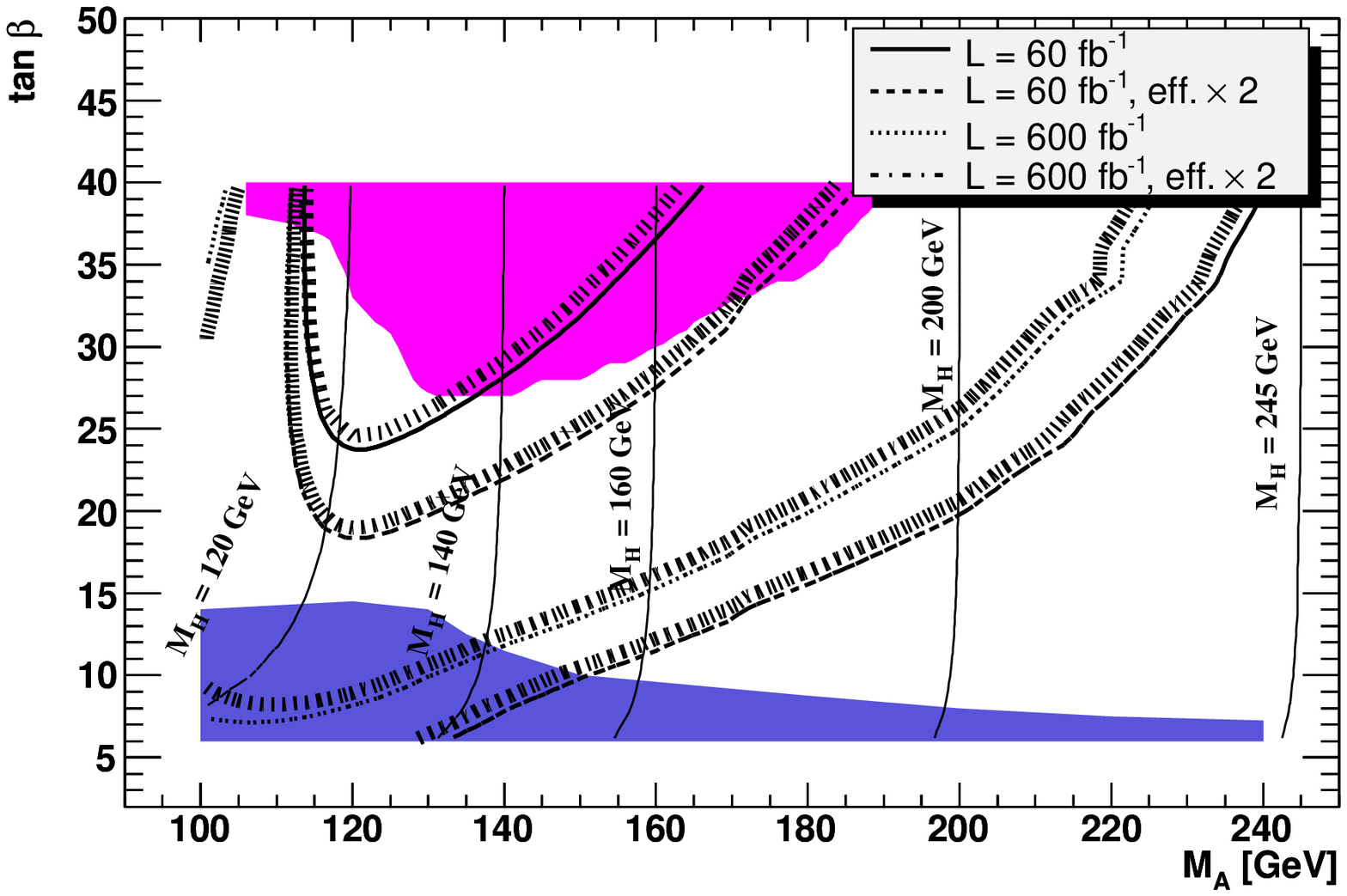}
\includegraphics[width=14cm,height=8.5cm]
                {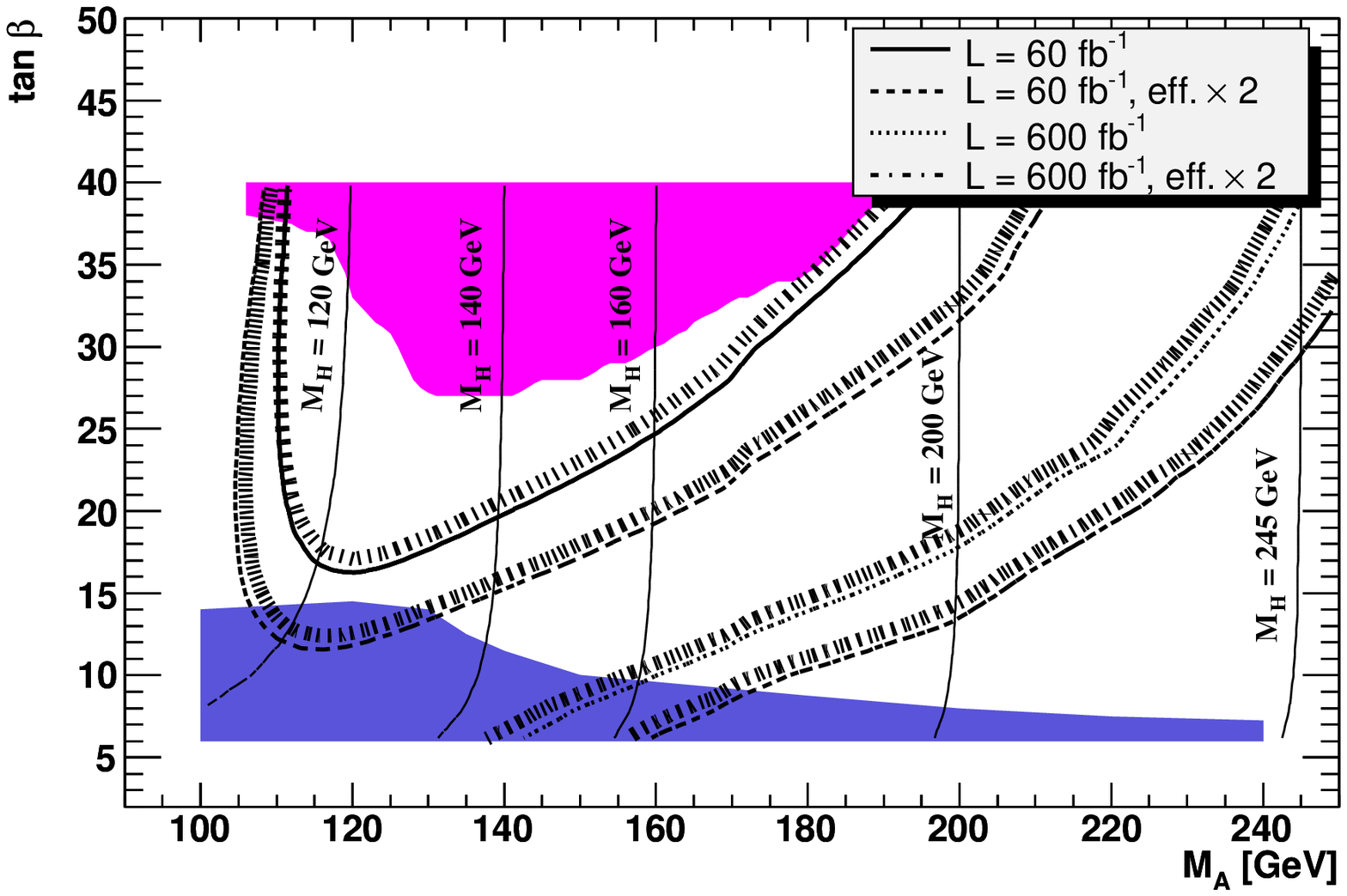}
%\vspace{1em}
\caption{
$5\,\si$ discovery contours (upper plot) and contours of $3\,\si$
statistical significance (lower plot) for the $H \to b \bar b$ channel in
CED production in the $\MA$--$\tb$ plane of the MSSM within the 
``CDM benchmark'' scenario \pvier. The results are shown for assumed
effective luminosities (see text, combining ATLAS and CMS) of \sixoo,
\sixooeff, \sixooo\ and \sixoooeff. 
The values of $\MH$ are
indicated by contour lines. The dark shaded (blue)
region corresponds to the parameter region that
is excluded by the LEP Higgs 
searches, the lighter shaded (pink) areas are excluded by Tevatron Higgs
searches. 
}
\label{fig:Hbb-p4}
\end{center}
\end{figure}

The corresponding results in the scenario \pvier\ are presented in
\reffi{fig:Hbb-p4}. The reach is similar to \pdrei\ (but restricted by
construction to $\tb \le 40$). Again, a large fraction of the parameter
space covered at $5\,\si$ with low luminosity is excluded by the Tevatron
searches, except for a region with $\MA \lsim 135 \gev$. The
$3\,\si$ significance contours, on the other hand, cover a large part of
the unexcluded region between the Tevatron and the LEP limits even for
the low-luminosity scenarios.
In the high-luminosity scenario at $\tb \lsim 40$ 
we find that $\MH = 240 (260) \gev$ can be covered at $5 (3)\,\si$.

%%%%%%%%%%%%%%%%%%%%%%%%%%%%%%%% Begin FIGURE %%%%%%%%%%%%%%%%%%%%%%%%%%%%%%%%%
%%%%%%%%%%%%%%%%%%%%%%%%%%%%%%%% End FIGURE %%%%%%%%%%%%%%%%%%%%%%%%%%%%%%%%%%%

%%%%%%%%%%%%%%%%%%%%%%%%%%%%%%%% Begin FIGURE %%%%%%%%%%%%%%%%%%%%%%%%%%%%%%%%%
%%%%%%%%%%%%%%%%%%%%%%%%%%%%%%%% End FIGURE %%%%%%%%%%%%%%%%%%%%%%%%%%%%%%%%%%%

%%%%%%%%%%%%%%%%%%%%%%%%%%%%%%%% Begin FIGURE %%%%%%%%%%%%%%%%%%%%%%%%%%%%%%%%%
\begin{figure}[htb!]
\vspace{0.5em}
\begin{center}
\includegraphics[width=14cm,height=8.5cm]
                {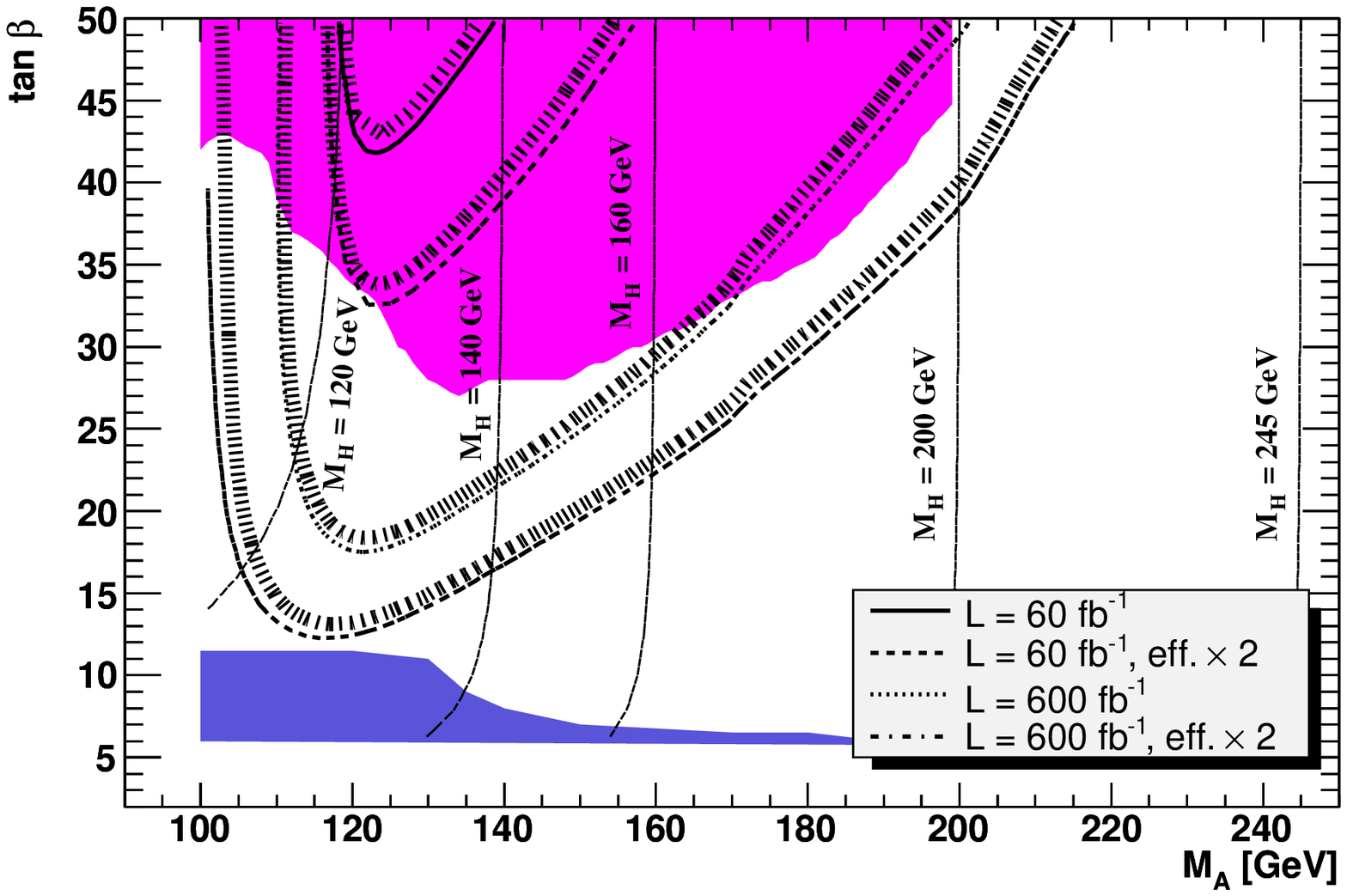}
\includegraphics[width=14cm,height=8.5cm]
                {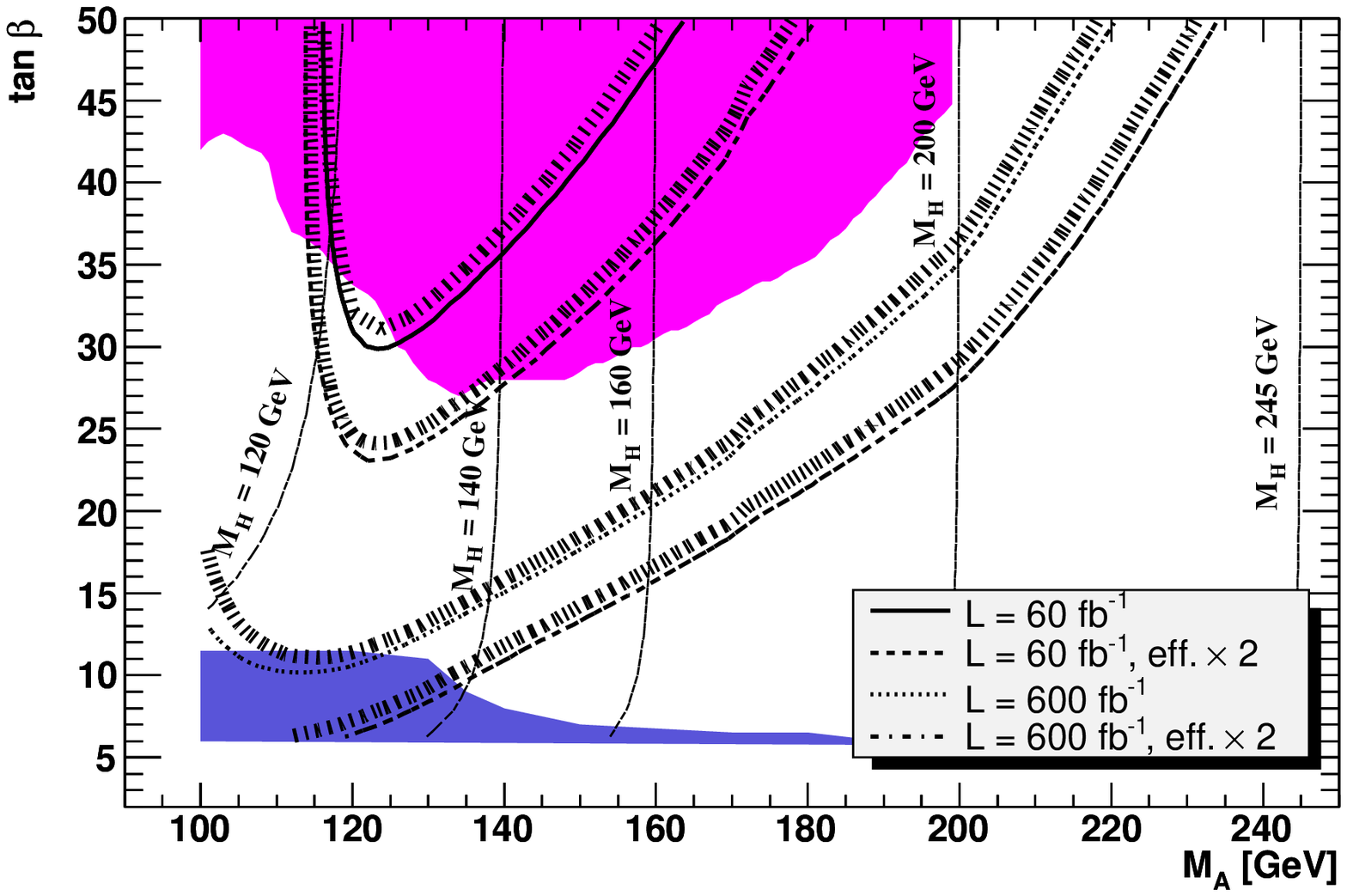}
%\vspace{1em}
\caption{
$5\,\si$ discovery contours (upper plot) and contours of $3\,\si$
statistical significance (lower plot) for the $H \to \tau^+\tau^-$ channel in
CED production in the $\MA$--$\tb$ plane of the MSSM within the 
``CDM benchmark'' scenario \pdrei. The results are shown for assumed
effective luminosities (see text, combining ATLAS and CMS) of \sixoo,
\sixooeff, \sixooo\ and \sixoooeff. 
The values of $\MH$ are
indicated by contour lines. The dark shaded (blue)
region corresponds to the parameter region that
is excluded by the LEP Higgs 
searches, the lighter shaded (pink) areas are excluded by Tevatron Higgs
searches. 
}
\label{fig:Htautau-p3}
\end{center}
\end{figure}
%%%%%%%%%%%%%%%%%%%%%%%%%%%%%%%% End FIGURE %%%%%%%%%%%%%%%%%%%%%%%%%%%%%%%%%%%

Finally we present the results for $H \to \tau^+\tau^-$ for the scenario
\pdrei\ in \reffi{fig:Htautau-p3}. The reach is slightly larger than in
the $\Mhmax$ scenario as shown in \reffi{fig:Htautau}. However, as in
the conventional benchmark scenario the regions covered at $5\,\si$
with low luminosity are excluded by the Tevatron searches.
The coverage in the 
\sixooo\ and \sixoooeff\ luminosity scenarios extends down to 
$\tb \lsim 15$ at $\MA \approx \MH \approx 120 \gev$, and at $\tb = 50$
to $\MA \lsim 210 \gev$.
The $3\,\si$ coverage at the highest luminosity
goes up to $\MA \approx 235 \gev$ at $\tb = 50$ 
and down to $\MA \approx 135 \gev$ around $\tb = 10$,
where the coverage of the CED channel overlaps with 
the LEP exclusion bound.

\smallskip
To summarize, 
from the investigation of the conventional scenarios and the ``CDM benchmark'' 
scenarios we can draw the conclusion that there are promising prospects 
for the 
CED production of the light and / or heavy $\cp$-even Higgs bosons of the
MSSM with subsequent decay to
$b \bar b$ (and possibly to $\tau^+\tau^-$).
The fact that we find similar results in the conventional
  benchmark scenarios and in the ``CDM benchmark'' scenarios
  strengthens the overall validity of our findings.
While in the \sixoo\ and \sixooeff\ scenarios
the results from the
Tevatron Higgs searches meanwhile exclude some of the $5\,\si$ regions
that are accessible for the heavy $\cp$-even Higgs in CED production, 
a significant coverage of currently unexplored parameter space 
can be achieved at the $3\,\si$ level. If the 
CED channel can be utilized at high instantaneous luminosity, as assumed
in the \sixooo\ and \sixoooeff\ scenarios, nearly the whole parameter
space in the $\MA$--$\tb$ plane can be covered at the $3\,\si$ level 
with the searches for CED production of 
the light and heavy neutral $\cp$-even Higgs bosons of the MSSM.
As discussed in more detail below,
the detection of a Higgs candidate in this channel would provide
important information about the properties of the observed state.

%%%%%%%%%%%%%%%%%%%%%%%%%%%%%%%%%%%%%%%%%%%%%%%%%%%%%%%%%%%%%%%%%%%%%%%%%%%%%%%
%%%%%%%%%%%%%%%%%%%%%%%%%%%%%%%%%%%%%%%%%%%%%%%%%%%%%%%%%%%%%%%%%%%%%%%%%%%%%%%

\section{CED Higgs production in a model with a fourth generation of
  fermions}
\label{sec:SM4}

A rather simple example of physics beyond the SM is a
model ``SM4'' which extends the SM by a fourth generation of heavy
fermions, see for instance \citeres{extra-gen-review,4G,4G-ew}. 
The masses of the 4th generation quarks in such a scenario need to be
significantly heavier than the mass of the top quark.%
\footnote{
The masses of the 4th generation leptons, which are essentially
irrelevant for our discussion below, are less restricted and could be
as light as about $100 \gev$.}%
~As a consequence, the effective coupling of the Higgs boson
to two gluons in the SM4 
is to good approximation three times larger than in the SM.
The effect of the 4th generation on the other couplings that are 
relevant for Higgs searches at LEP and the Tevatron is small. 
The phenomenological impact on the Higgs searches can therefore 
be described by a change of the partial decay width $\Gamma(H\to gg)$ 
by a factor of 9, giving rise to a corresponding shift in the
total Higgs width and therefore all the decay branching ratios, see for
instance \citere{four-gen-and-Higgs,ggH4}.
The total decay width in the SM4 and the relevant decay branching 
ratios can to good approximation 
be evaluated in terms of the corresponding quantities in the SM as
\newcommand{\TOT}{{\rm tot}}
\newcommand{\SMv}{{\rm SM4}}
\newcommand{\HSMv}{H^{\SMv}}
\newcommand{\MHSMv}{M_{\HSMv}}
\begin{align}
\Gamma_\SM(H\to gg) &= \br_\SM(H\to gg)\:\Gamma_\TOT^\SM(H)\,,\\ 
\Gamma_\SMv(H\to gg) &= 9\:\Gamma_\SM(H\to gg)\,,\\
\Gamma_\TOT^\SMv(H) &= \Gamma_\TOT^\SM(H) - \Gamma_\SM(H\to gg)
	+ \Gamma_\SMv(H\to gg)\,.
\end{align}
Because of the enhanced effective coupling of the Higgs boson to two
gluons in the SM4, the Tevatron Higgs searches based in particular on 
the channel $gg \to H \to WW^{(*)}$ have a significantly higher
sensitivity as compared to the SM case. In a recent combined analysis 
of the CDF and D\O collaborations the mass range 
$130 \gev \lsim \MHSMv \lsim 210 \gev$ could be excluded at the 
95\% C.L.~\cite{SM4-CDF-D0}. The LEP Higgs searches, on the other hand, 
exclude the Higgs boson of the SM4 for Higgs masses below $\sim 112 \gev$,
see the discussion in \citere{higgsbounds}. Thus, there remains an
unexcluded mass window for a relatively light Higgs in the SM4 
of $112 \gev \lsim \MHSMv \lsim 130 \gev$. 
We now investigate the sensitivity of CED Higgs production at the LHC 
to the Higgs boson of the fourth generation model within this mass window.

%%%%%%%%%%%%%%%%%% F I G U R E %%%%%%%%%%%%%%%%%%%%%%%%%%%%%%%%%%%%%%%%%%%%%%
\begin{figure}[htb!]
%\vspace{1em}
\begin{center}
\includegraphics[width=14cm,height=8.7cm]{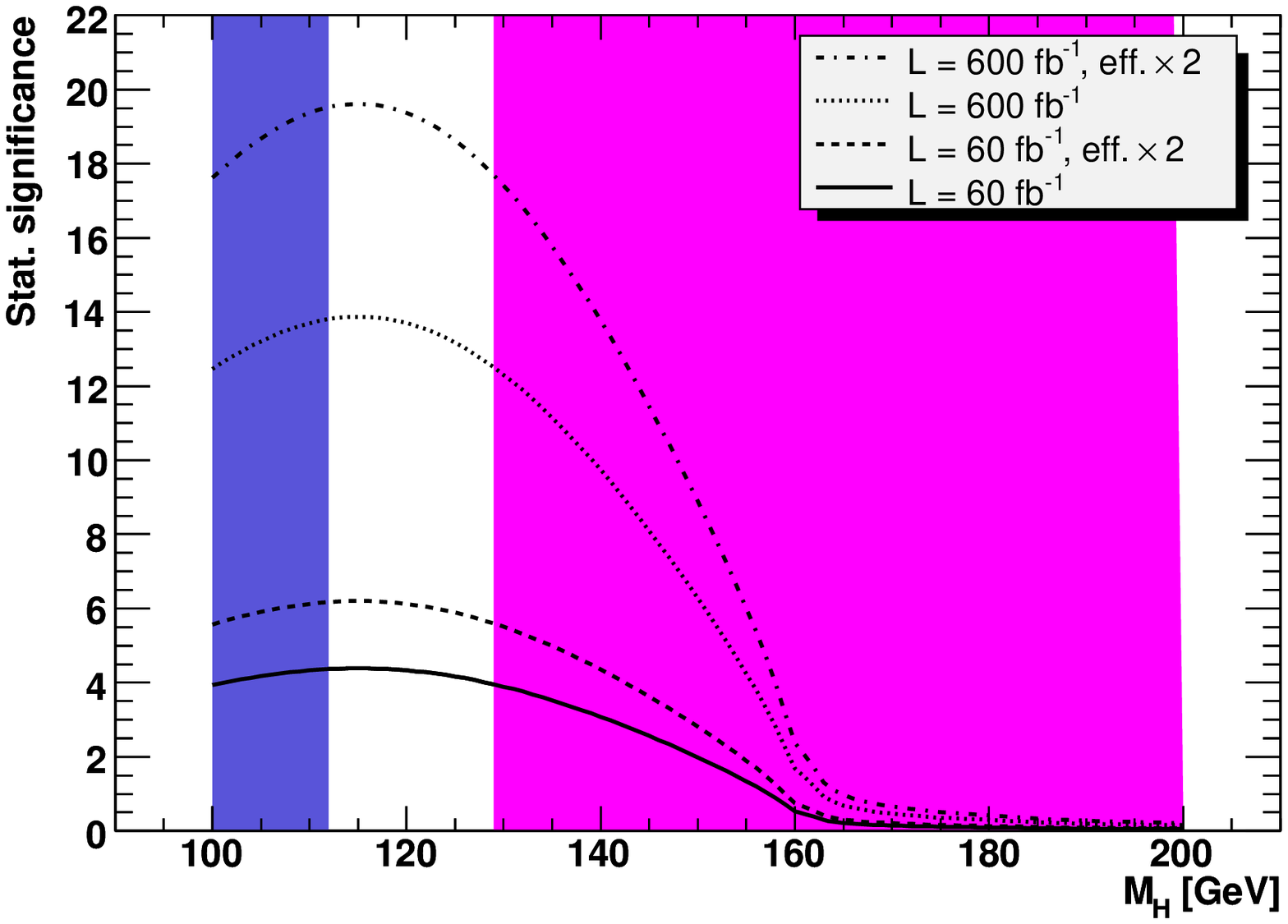}\\[1em]
\includegraphics[width=14cm,height=8.7cm]{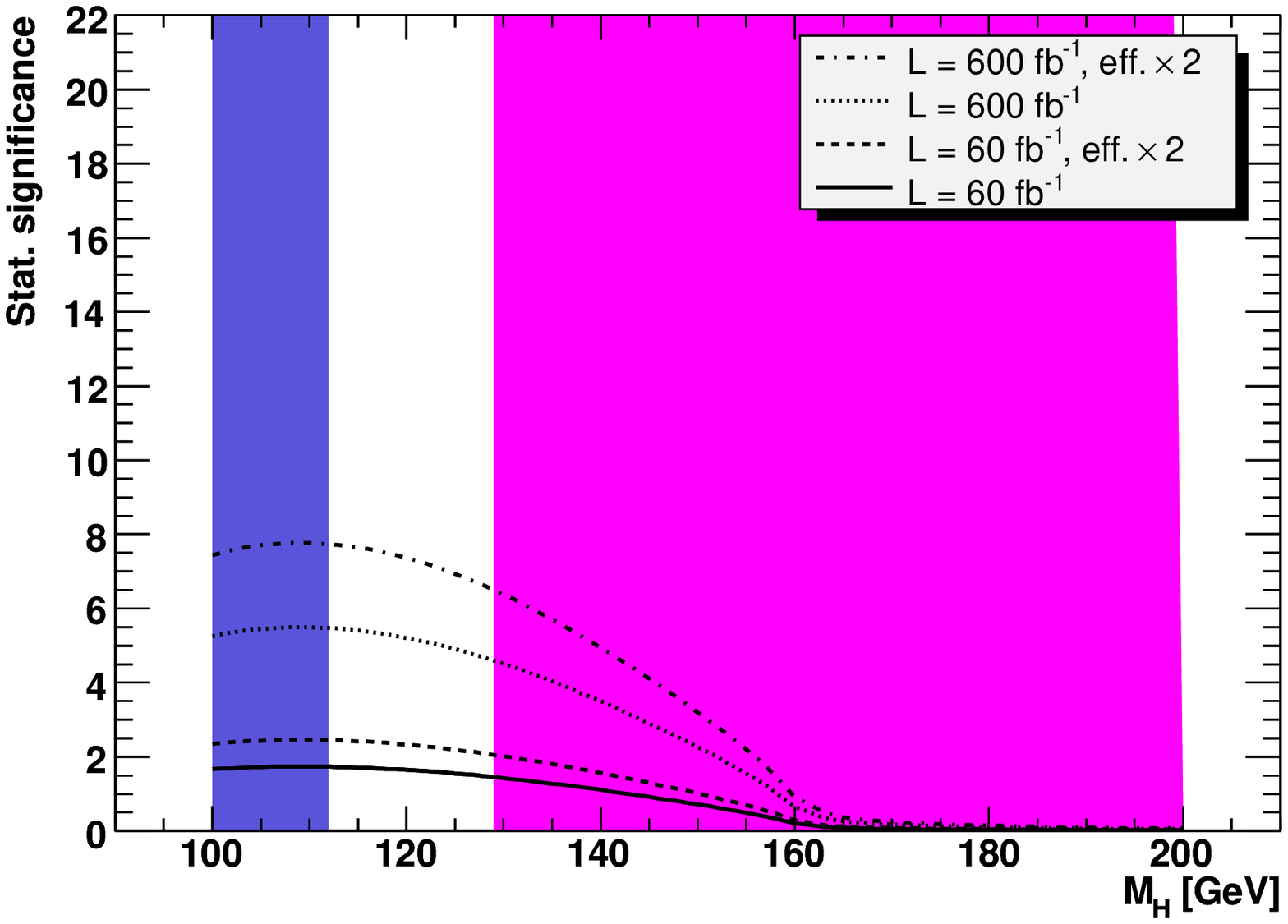}
\vspace{1em}
\caption{Significances reachable with CED Higgs production in the SM4
  in the $H \to b \bar b$ 
  (upper plot) and $H \to \tau^+\tau^-$ (lower plot) channel 
for effective luminosities of ``\sixoo'',
``\sixooeff'', ``\sixooo'' and ``\sixoooeff''.
The regions excluded by LEP appear as blue/light grey for low values of 
$\MHSMv$ and those 
excluded by the Tevatron as red/dark grey for larger values of
$\MHSMv$. 
}
\label{fig:disc-SM4}
\end{center}
%\vspace{1em}
\end{figure}
%%%%%%%%%%%%%%%%%% F I G U R E %%%%%%%%%%%%%%%%%%%%%%%%%%%%%%%%%%%%%%%%%%%%%%

As for our analysis within the MSSM discussed above, 
we have evaluated the significances that can be obtained
in the channels $H \to b \bar b$ and $H \to \tau^+\tau^-$. The results
are shown in \reffi{fig:disc-SM4} as a function of $\MHSMv$ for the four
luminosity scenarios discussed above. 
The regions excluded by LEP appear as blue/light grey for low values of 
$\MHSMv$, and regions excluded by the Tevatron appear as red/dark grey
for larger values of $\MHSMv$. 
The $b \bar b$ channel (upper plot) shows that even at rather low
luminosity the remaining window of $112 \gev \lsim \MHSMv \lsim 130 \gev$
can be covered by CED Higgs production. Due to the smallness of 
$\br(\HSMv \to b \bar b)$ at $\MHSMv \gsim 160 \gev$, however, this
channel becomes irrelevant for the still allowed high values of
$\MHSMv$, and we do not extend our analysis beyond 
$\MHSMv \le 200 \gev$. The $\tau^+\tau^-$ channel (lower plot) 
reaches a sensitivity of about $2\,\si$ at low
luminosity, while it can exceed the $5\,\si$ level at high LHC luminosity.
At masses $\MHSMv \gsim 220 \gev$ it might be possible to exploit the decay 
$H \to WW, ZZ$, but (apart from the recent analysis in
\citere{SM4spannowsky}) no detailed investigation has been performed up
to now.

%%%%%%%%%%%%%%%%%%%%%%%%%%%%%%%%%%%%%%%%%%%%%%%%%%%%%%%%%%%%%%%%%%%%%%%%%%%%%%%
%%%%%%%%%%%%%%%%%%%%%%%%%%%%%%%%%%%%%%%%%%%%%%%%%%%%%%%%%%%%%%%%%%%%%%%%%%%%%%%

\section{Spin--parity and coupling structure determination of Higgs
  bosons at the LHC} 
\label{sec:spin}

The standard methods for determining the spin and the $\cp$ properties
of Higgs bosons at the LHC rely to a large extent on the
coupling of a relatively heavy Higgs boson to two gauge bosons.
In particular, the channel $H \to ZZ \to 4l$
-- if it is open -- offers good prospects 
for studying spin and $\cp$-properties of
possible Higgs candidates~\cite{jakobs-rev} (see also
\citere{Hlookalike} for a recent analysis of this channel in the context
of distinguishing a SM Higgs boson from other possible states of new
physics).

Furthermore, the weak boson fusion channel at the LHC has been 
investigated as a means to obtain information on the coupling of a
Higgs-like state to two gauge bosons~\cite{HVV-LHC0,HVV-LHC2,HVV-LHC1}.
In a study exploiting the difference in the
azimuthal angles of the two tagging jets in weak vector boson 
fusion~\cite{HVV-LHC1} it was found that for $\MHSM = 160 \gev$ 
the decay mode into a pair of $W$-bosons (for which the branching ratio 
is maximal in this mass range)
allows the discrimination between the two extreme
scenarios of a pure $\cp$-even (as in the SM) and a pure $\cp$-odd
tensor structure at a level of 4.5--5.3\,$\si$ using about 10~\ifb 
of data
(assuming the production rate is that of the SM, which is in conflict
with the latest search limits from the Tevatron~\cite{TevHiggsSM}.)
A discriminating power of two standard deviations at $\MHSM = 120 \gev$
in the tau lepton decay mode requires an integrated luminosity of
30~\ifb~\cite{HVV-LHC1}. 

For $\MH \approx \MA \gsim 2 \MW$ the lightest MSSM
Higgs boson couples to gauge bosons with about SM strength, but its mass
is bounded from above by $\Mh \lsim 135 \gev$~\cite{mhiggsAEC},
i.e.\ the light Higgs is in a mass range where the decay to
$WW^{(*)}$ or $ZZ^{(*)}$ is difficult to exploit.
On the other hand, the 
heavy MSSM Higgs bosons, $H$ and $A$, decouple from the gauge bosons,
see \refse{sec:mssmhiggs}. 
Consequently, since the usually quoted results for the 
$H \to ZZ \to 4l$ channel and the weak-boson fusion channel with 
decay into a pair of $W$ bosons assume a relatively 
heavy ($\MH \gsim 135 \gev$) SM-like Higgs, these results are not
applicable to the case of the MSSM. The same is true for any other model 
with a light SM-like Higgs and heavy Higgs bosons that decouple from
the gauge bosons (it should be noted in this context that in certain
parameter regions of general Two-Higgs-Doublet-Models a sizable
branching fraction of a heavy scalar or pseudo-scalar Higgs into a pair
of gauge bosons is possible~\citere{HVV-2HDM}). On the other hand, the 
analysis mentioned above in the
weak boson fusion channel with a subsequent decay into a pair of $\tau$
leptons can be utilised for a relatively light SM-like Higgs and is
therefore applicable also to the light $\cp$-even Higgs boson in the
MSSM. However, in the MSSM 
no significant enhancement of this channel would be expected compared to
the SM case, so that the rather modest sensitivities found for the
studies in the SM would also hold for the case of the MSSM.

Accordingly, alternative methods for determining the spin and 
$\cp$-properties of Higgs boson candidates that do not rely on the decay
into a pair of gauge bosons or on the production in weak boson fusion
are of great phenomenological interest. The CED Higgs-boson production
process can yield crucial information in this 
context~\cite{KMRProsp,KKMRext,diffH,eds09}. 
Because of the $J_z=0$, $C$-even, $P$-even selection rule, which holds for all 
CED processes, the fact that a certain state is produced in a CED process 
provides direct information about its spin and $\cp$-properties.
It is expected, in particular in a situation where a new particle state
has also been detected in one or more of the conventional Higgs search
channels, that already a small yield of CED events will be sufficient
for extracting relevant information on the spin and $\cp$-properties of
the new state.

It should be noted here that most of the existing analyses and also the
discussion in the present paper have been performed in the context of
the $\cp$-conserving MSSM, i.e.\ in terms of the $\cp$
eigenstates $h, H, A$. The CED Higgs-boson analyses 
can easily be extended to the case 
where $\cp$-violating complex phases are present, giving rise to a 
mixing between the three neutral Higgs bosons. In the general case where
the produced state is not necessarily a $\cp$-eigenstate, the selection
rule of the CED process can be interpreted as performing a projection of
a possibly mixed state onto a $\cp$-eigenstate.

It is worth mentioning that measuring the transverse momentum
and azimuthal angle $\phi$ distributions of forward  leading protons will
provide a unique opportunity to search for a $\cp$-violating signal
in the Higgs sector~\cite{CP}.
In particular in some MSSM scenarios with $\cp$-violation
the azimuthal asymmetry of the outgoing tagged protons
\begin{align}
A &= \frac{\si(\varphi<\pi)-\si(\varphi>\pi)}
          {\si(\varphi<\pi)+\si(\varphi>\pi)}
\end{align}
is expected to be quite sizeable. For instance~\cite{CP,ACCOM},  
$A \simeq 0.07$ can be reached in the scenarios of \citeres{je2,CEPW}.

As discussed in \citeres{KKMRext,diffH} it will be challenging to identify
the  $\cp$-odd Higgs boson of the MSSM, $A$, in the CED processes
because of the strong suppression, caused by the
$P$-even selection rule, which effectively filters out its production.
However, in the semi-inclusive diffractive reactions the pseudoscalar
production is much less suppressed. A recent study in \citere{KMRbsm} shows
that there are certain advantages of looking for the $\cp$-odd Higgs particle
in the semi-inclusive process $pp \to p + gAg + p$
with  two tagged forward  protons and two large rapidity gaps.
In this case the amplitude of $\cp$-odd $A$~boson
production is suppressed in comparison with the $\cp$-even boson amplitude
only by the momentum fraction $x$ carried by the accompanying
gluon. Thus, in the events with one relatively hard gluon, whose energy
is comparable with the energy of the whole $g A g$ system, 
the cross sections of $\cp$-odd $A$~boson production will be of the same
order as that for the $\cp$-even boson.

%%%%%%%%%%%%%%%%%%%%%%%%%%%%%%%%%%%%%%%%%%%%%%%%%%%%%%%%%%%%%%%%%%%%%%%%%%%%%%%
%%%%%%%%%%%%%%%%%%%%%%%%%%%%%%%%%%%%%%%%%%%%%%%%%%%%%%%%%%%%%%%%%%%%%%%%%%%%%%%

\section{Conclusions}
\label{sec:conclusions}

We have analysed in this paper the prospects for probing the Higgs
sector of models beyond the SM with central exclusive Higgs-boson production
processes at the LHC, utilizing forward proton detectors installed
at 220~m and 420~m distance around ATLAS and CMS.
We have updated previous results on the prospects 
for CED production of the neutral $\cp$-even
Higgs bosons $h$ and $H$ 
of the MSSM and their decays into bottom quarks and $\tau$
leptons. 
With respect to \citere{diffH} an improved cross section calculation 
and background evaluation has been performed. Furthermore, the updated
exclusion regions from Higgs-boson searches at the Tevatron have
been taken into account (and the most complete compilation of the LEP
exclusion bounds has been used).
In the conventional benchmark scenarios, the $\Mhmax$ and the no-mixing
scenario, we find that the 
results show a larger coverage in the $\MA$--$\tb$ parameter space in
comparison with \citere{diffH}. On the other hand, the latest results
from the Higgs searches at the Tevatron also yield an enlargement of
the excluded regions in the $\MA$--$\tb$, which rule out parts of the
parameter regions where the CED Higgs production channels at the LHC 
have a sensitivity in excess of $5\,\si$. 

We have further extended the results previously obtained for the MSSM 
by investigating additional ``CDM benchmark'' scenarios, i.e.\
$\MA$--$\tb$ planes that are in agreement
with the relic abundance of cold dark matter as well as with electroweak
precision observables and $B$~physics. As a qualitative result, valid
for all analysed benchmark scenarios, we find that at the LHC 
the CED production channels of the light and heavy $\cp$-even Higgs
bosons of the MSSM with subsequent decay to
$b \bar b$ and, to a lesser extent, to $\tau^+\tau^-$ provide promising
sensitivity.
While in the considered scenarios with lower instantaneous luminosity,
\sixoo\ and \sixooeff, some of the $5\,\si$ regions accessible for the
heavy $\cp$-even Higgs in CED production are excluded by the Tevatron
Higgs searches, 
a significant coverage of currently unexplored parameter space 
can be achieved at the $3\,\si$ level. For the \sixooo\ and \sixoooeff\
scenarios, which could be realized if the
CED channel can be utilized at high instantaneous luminosity, we have
demonstrated that the observation of either the light or the heavy 
$\cp$-even Higgs boson of the MSSM becomes possible at the $3\,\si$
level in large parts of the MSSM parameter space.

A striking feature of CED Higgs-boson production remains that this channel 
provides good prospects for detecting Higgs-boson decays into bottom
quarks and $\tau$ leptons (and also into $W$ bosons, which has not been
updated here). Although the decay into bottom
quarks is the dominant decay mode for a light SM-like Higgs boson, this
decay channel is difficult to access in the conventional search
channels at the LHC. In the MSSM the $b \bar b$ and $\tau^+\tau^-$ decay
channels are of particular importance, since they are in general 
dominant even for heavy MSSM Higgs bosons, whereas a SM Higgs boson of the
same mass would have a negligible branching ratio into $b \bar b$ and
$\tau^+\tau^-$.
It should be noted that 
heavy Higgs bosons that decouple from gauge bosons and therefore
predominantly decay into heavy SM fermions (or into other states of new
physics, if decays of this kind are kinematically possible)
are a quite generic feature
of extended Higgs-boson sectors. 

As another example of Higgs physics beyond the SM we have analysed the
prospects of CED Higgs production in the SM4, i.e.\ the SM with a fourth
generation of (heavy) fermions. 
In the Higgs mass range below about $200 \gev$, the
LEP and Tevatron Higgs-boson searches
leave only a relatively small unexcluded window for the SM4 Higgs-boson
mass, $112 \gev \lsim \MHSMv \lsim 130 \gev$. We have shown that within
this window the CED Higgs production at the LHC 
with the subsequent decays
$\HSMv \to b \bar b, \tau^+\tau^-$ offers very good prospects for
analysing the Higgs sector of the model. For $\MHSMv \gsim 200 \gev$ the
decay to SM gauge bosons becomes dominant, rendering the decay to 
(light) SM fermions ineffective.

Finally we have discussed the prospects for determining the spin 
and the coupling structure of possible Higgs candidates at the LHC. The
existing analyses assuming a SM-like Higgs boson 
can only be partially applied to the case of the light $\cp$-even
Higgs boson of the MSSM and cannot be employed for the heavy $\cp$-even
Higgs boson of the MSSM nor for any other extension of the SM with a 
light SM-like Higgs and heavy Higgs bosons that decouple from the gauge
bosons. We emphasize that the $J_z = 0$, $\cC$-even, $\cP$-even
selection rule of the CED Higgs production process 
leads to a direct
information on the properties of any particle produced in this
way. 

While for clarity of presentation, the results discussed in this paper
have been phrased in terms of the $\cp$ eigenstates $h, H, A$ of the Higgs
bosons, a generalisation to the general case of arbitrary $\cp$-violating
mixing between the neutral Higgs bosons is easily possible. In this case
the selection rule of the CED process has the effect of performing a
projection of a possibly mixed state onto a $\cp$ eigenstate. 
Accordingly, the production of a $\cp$-odd boson in the CED processes is
strongly suppressed. Access to $\cp$-odd Higgs production could
potentially be obtained in semi-inclusive processes. However, further 
work will
be needed to assess the viability of such production modes.

%%%%%%%%%%%%%%%%%%%%%%%%%%%%%%%%%%%%%%%%%%%%%%%%%%%%%%%%%%%%%%%%%%%%
%%%%%%%%%%%%%%%%%%%%%%%%%%%%%%%%%%%%%%%%%%%%%%%%%%%%%%%%%%%%%%%%%%%%

\subsection*{Acknowledgements}

We thank  Mike Albrow, Oliver Brein, Peter Bussey, Albert De Roeck,
John Ellis, David d'Enterria, 
Alan Martin, Risto Orava, Krzysztof Piotrzkowski, Christophe Royon, Andy 
Pilkington, James~Stirling and Tim Tait for useful discussions.
MGR thanks the IPPP at the University of Durham for hospitality. 
The work by MGR was supported by the Federal Program of the Russian State
RSGSS-3628.2008.2. 
The work of MT was supported by the 
project AV0-Z10100502 of the Academy of Sciences of the Czech republic and
project LC527 of the Ministry of Education of the Czech republic.
The work of SH was partially supported by CICYT (grant FPA 2007--66387).
This work is also supported in part by the European Community's
Marie-Curie Research Training Network under contract MRTN-CT-2006-035505
`Tools and Precision Calculations for Physics Discoveries at Colliders'.

%%%%%%%%%%%%%%%%%%%%%%%%%%%%%%%%%%%%%%%%%%%%%%%%%%%%%%%%%%%%%%%%%%%%
%%%%%%%%%%%%%%%%%%%%%%%%%%%%%%%%%%%%%%%%%%%%%%%%%%%%%%%%%%%%%%%%%%%%

%%%%%%%%%%%%%%%%%%%%%%%%%%%%%%%%%%%%%%%%%%%%%%%%%%%%%%%%%%%%%%%%%%%%%%%%%%%%%%%
%%%%%%%%%%%%%%%%%%%%%%%%%%%%%%%%%%%%%%%%%%%%%%%%%%%%%%%%%%%%%%%%%%%%%%%%%%%%%%%

%\newpage

\end{document}

%xxxxxxxxxxxx

%%%%%%%%%%%%%%%%%%%%%%%%%%%%%%%%%%%%%%%%%%%%%%%%%%%%%%%%%%%%%%%%%%%%%%%%%%%%%%%
%%%%%%%%%%%%%%%%%%%%%%%%%%%%%%%%%%%%%%%%%%%%%%%%%%%%%%%%%%%%%%%%%%%%%%%%%%%%%%%